\documentclass[12pt]{article}

\usepackage{amsmath}
\usepackage{amssymb}
\usepackage{epsfig}
\usepackage{graphicx}
\usepackage{cite}
\newcommand{\capdef}{}
\newcommand{\mycaption}[2][\capdef]{\renewcommand{\capdef}{#2}%
        \caption[#1]{{\itshape #2}}}
\makeatletter
\renewcommand{\fnum@table}{\textbf{\tablename~\thetable}}
\renewcommand{\fnum@figure}{\textbf{\figurename~\thefigure}}
\makeatother

\newcounter{myenumi}

\renewcommand{\themyenumi}{\roman{myenumi}}
{\end{list}}

\newlength{\myem}
\newcounter{mysubequation}[equation]

\makeatletter
\renewcommand{\section}{\@startsection{section}{1}{0em}{-\baselineskip}%
{\baselineskip}{\normalfont\large\bfseries}}
\renewcommand{\subsection}%
{\@startsection{subsection}{2}{0em}{-0.7\baselineskip}%
{0.7\baselineskip}{\normalfont\bfseries}}
\makeatother

\newcommand{\bea}{\begin{eqnarray*}}
\newcommand{\eea}{\end{eqnarray*}}

\newcommand{\deltacp}{\delta_\mathrm{CP}}

\newcommand{\GeV}{\,\mathrm{GeV}}

\newcommand{\eV}{\,\mathrm{eV}}

\newcommand{\ree}{{\nu_e\rightarrow\nu_e}}
\newcommand{\rue}{{\nu_\mu\rightarrow\nu_e}}

\newcommand{\ruu}{{\nu_\mu\rightarrow\nu_\mu}}
\newcommand{\rux}{{\nu_\mu\rightarrow\nu_x}}

\newcommand{\reeb}{{\bar{\nu}_e\rightarrow\bar{\nu}_e}}

\newcommand{\dm}[1]{{\Delta m^2_{#1}}}

\newcommand{\ie}{{\it i.e.}}

\newcommand{\cf}{{\it cf.}}

\newcommand{\eq}{Equation}
\newcommand{\eqs}{Equations}

\newcommand{\fig}{Figure}

\newcommand{\Ref}{Ref.}
\newcommand{\Refs}{Refs.}
\newcommand{\Sec}{Section}

\newcommand{\App}{the Appendix}

\newcommand{\Tab}{Table}

\begin{document}
%%%%%%%%%%%%%%%%%%%%%%%%%%%%%%%%%%%%%%%%%%%%%%%%%%%%%%%%%%%%%%%%%%%%%
%%%%                     Title-page                              %%%%
%%%%%%%%%%%%%%%%%%%%%%%%%%%%%%%%%%%%%%%%%%%%%%%%%%%%%%%%%%%%%%%%%%%%%

\begin{titlepage}

% the footnote symbols are only redefined for the title page !
\renewcommand{\thefootnote}{\alph{footnote}}

\vspace*{-3.cm}
\begin{flushright}
TUM-HEP-464/02\\
MPI-PhT/2002-69\\
%hep-ph/yymmnnn
\end{flushright}

\vspace*{0.5cm}

\renewcommand{\thefootnote}{\fnsymbol{footnote}}
\setcounter{footnote}{-1}

{\begin{center}
{\large\bf Synergies between the first-generation JHF-SK and NuMI superbeam 
experiments$^*$\footnote{\hspace*{-1.6mm}$^*$Work supported by
``Sonderforschungsbereich 375 f\"ur Astro-Teilchenphysik'' der Deutschen
Forschungsgemeinschaft and the ``Studienstiftung des deutschen Volkes'' (German
National Merit Foundation) [W.W.].}} \end{center}}
\renewcommand{\thefootnote}{\alph{footnote}}

\vspace*{.8cm}
\vspace*{.3cm}
{\begin{center} {\large{\sc
                P.~Huber\footnote[1]{\makebox[1.cm]{Email:}
                phuber@ph.tum.de},~
                M.~Lindner\footnote[2]{\makebox[1.cm]{Email:}
                lindner@ph.tum.de},~and~
                W.~Winter\footnote[3]{\makebox[1.cm]{Email:}
                wwinter@ph.tum.de}~
                }}
\end{center}}
\vspace*{0cm}
{\it
\begin{center}

\footnotemark[1]${}^,$\footnotemark[2]${}^,$\footnotemark[3]%
       Institut f\"ur theoretische Physik, Physik--Department,
       Technische Universit\"at M\"unchen,\\
       James--Franck--Strasse, D--85748 Garching, Germany

\footnotemark[1]%
       Max-Planck-Institut f\"ur Physik, Postfach 401212,
       D--80805 M\"unchen, Germany

\end{center}}

\vspace*{1.5cm}

{\Large \bf
\begin{center} Abstract \end{center}  }

We discuss synergies in the combination of the first-generation JHF to 
Super-Kamiokande and NuMI off-axis superbeam
experiments. With synergies we mean effects which go 
beyond simply adding the statistics of the
two experiments. As a first important result,
we do not observe interesting synergy effects in the combination of 
the two experiments as they are planned right now.
However, we find that with minor modifications, such as a different
NuMI baseline or a partial antineutrino running, one could do much richer 
physics with both experiments combined. Specifically, we demonstrate that
one could, depending on the value of the solar mass squared difference,
either measure the sign of the atmospheric mass squared difference or CP violation
already with the initial stage experiments. Our main results are presented in
a way that can be easily interpreted in terms of the forthcoming KamLAND result.

\vspace*{.5cm}

\end{titlepage}

\newpage

\renewcommand{\thefootnote}{\arabic{footnote}}
\setcounter{footnote}{0}

%%%%%%%%%%%%%%%%%%%%%%%%%%%%%%%%%%%%%%%%%%%%%%%%%%%%%%%%%%%%%%%%%%%%%
%                     Introduction                                  %
%%%%%%%%%%%%%%%%%%%%%%%%%%%%%%%%%%%%%%%%%%%%%%%%%%%%%%%%%%%%%%%%%%%%%

\section{Introduction}

There exists now very strong evidence for atmospheric neutrino 
oscillations, since there is some sensitivity to the characteristic 
$L/E$ dependence of oscillations~\cite{Toshito:2001dk}. Solar neutrinos
also undergo flavor transitions~\cite{Ahmad:2002jz,Ahmad:2002ka} solving
the long standing solar neutrino problem, even though the $L/E$ dependence of 
oscillations is in this case not yet established. Oscillations are,
however, by far the most plausible explanation and global fits to all 
available data clearly favor the so-called LMA solution for the mass 
splittings and mixings~\cite{Barger:2002iv,Bandyopadhyay:2002xj,%
Bahcall:2002hv,deHolanda:2002pp}. The CHOOZ reactor 
experiment~\cite{Apollonio:1999ae} moreover currently provides the most stringent 
upper bound for the sub-leading $U_{e3}$ element of the neutrino mixing matrix.
The atmospheric $\dm{31}$-value defines via $\dm{31}L/E_\nu=\pi/2$ the 
position of the first oscillation maximum $L$ as function 
of the energy $E_\nu$.
This leads for typical energies of $E_{\nu}\sim1 - 100 \, \mathrm{GeV}$ to
long-baseline (LBL) experiments, such as the ongoing
 K2K experiment~\cite{Nakamura:2001tr}, the 
MINOS~\cite{Paolone:2001am} and CNGS~\cite{Ereditato:2001an} experiments just being
constructed, and planned ``superbeam''~\cite{Barger:2000nf,%
Gomez-Cadenas:2001eu,Itow:2001ee,Minakata:2001qm,Aoki:2001rc,Aoki:2002ks,Barenboim:2002zx,Whisnant:2002fx,Aoki:2002ae,Barenboim:2002nv,Ayres:2002nm} or 
``neutrino factory'' experiments~\cite{DeRujula:1998hd,Barger:1999jj,FLPR,%
CERVERA,Barger:2000yn,FHL,Albright:2000xi,Burguet-Castell:2001ez,Freund:2001ui,%
Yasuda:2001ip,Bueno:2001jd,Yasuda:2002jk,Apollonio:2002en,Donini:2002rm}.
The solar $\dm{21}$ is, for the favored LMA solution, about two orders of 
magnitude smaller than the atmospheric $\dm{31}$, resulting in much longer
oscillations lengths for similar energies.
The solar oscillations will thus not fully develop in LBL experiments, 
but sub-leading effects play an important role in precision 
experiments. Together with matter effects~\cite{Wolfenstein:1978ue,%
Wolfenstein:1979ni,Mikheev:1985gs,Mikheev:1986wj}, interesting physics
opportunities are, in principle, opened, such as the extraction of the 
sign of $\dm{31}$ or the detection of CP violation in the lepton sector.
However, correlations and intrinsic 
degeneracies~\cite{Minakata:2001qm,Burguet-Castell:2001ez,Barger:2001yr} 
in the neutrino oscillation formulas turn
out to make the analysis of LBL experiments very complicated. Fortunately,
the potential to at least partially resolve those by the combination of
different experiments has been discovered~\cite{Barger:2002rr,Minakata:2002qi,%
Burguet-Castell:2002qx,Whisnant:2002fx,Barger:2002xk,Minakata:2002jv}.

Experimental LBL studies often assume that experiments
with different levels of sensitivity are built successively after each other.
For example, a frequent scenario is that a superbeam experiment (especially JHF) 
is followed by a neutrino factory with one or several baselines. 
However, there exist competing superbeam proposals, such as for the
the JHF to Super-Kamiokande ({\sf JHF-SK})~\cite{Itow:2001ee} and the
{\sf NuMI}~\cite{Ayres:2002nm} projects, which are some of the most promising
alternatives for the next generation LBL experiments.
Both are, in their current form,
optimized for themselves for a similar value of $L/E$ corresponding to the current 
atmospheric best-fit value of $\Delta m_{31}^2 = 3.0 \cdot 10^{-3}\, \mathrm{eV}^2$.
Building both experiments together would lead
to improved measurements because of the luminosity
increase, but one may ask if there are synergies in this
combination, \ie, complementary effects which not only come from 
the better statistics of both experiments combined. 
In this paper, we study, from a purely 
scientific point of view, the question if the two experiments 
are more than the sum of their parts, or if it would (in principle) 
be better to combine resources into one larger project. We are especially interested
in comparable versions of the initial stage options with running times of five
years each, which means that the first results could be obtained within this
decade. Thus, in comparison to \Ref~\cite{Barger:2002xk}, we do not assume
high-luminosity upgrades of both superbeam experiments, which leads to  somewhat 
different effects in the analysis because we are operating in the (low) statistics dominated
regime. In addition, we include in our analysis systematics, multi-parameter correlations, 
external input, such as from the KamLAND experiment, degeneracies, and the
matter density uncertainty.

In \Sec~\ref{sec:synergies}, we will define synergy effects in more detail. Next, we will address in \Sec~\ref{sec:experiments} (together with some more details in \App)
the experiments and their simulation, as well as we will show in \Sec~\ref{sec:framework}
the relevant analytical structure of the neutrino
oscillation framework. As the first part of the results, we demonstrate in \Sec~\ref{sec:initialstage}
that there are no real synergies between the experiments as they are planned right now. 
As the second part, we will investigate in \Sec~\ref{sec:alternatives}  alternative option to improve the physical
potential of the combined experiments. Eventually, we will summarize in 
\Sec~\ref{sec:summary} the best of those alternative options
together with a possible strategy to proceed. 

\section{How to define synergy effects}
\label{sec:synergies}

We define the extra gain in the combination of the experiments
beyond a simple addition of statistics as the synergy among two or more 
experiments. A reasonable definition of
synergy must therefore subtract, in a suitable 
way, the increase of statistics of otherwise more or less equivalent 
experiments. In this section, we demonstrate that one possibility 
to evaluate the synergy effects is to compare the output of the combination of
experiments, \ie, the sensitivity or precision of certain observables, 
to a reference experiment with the corresponding scaling of the integrated luminosity.

A LBL experiment is, beyond statistics, influenced by
several different sources of errors, such as systematical errors, multi-parameter correlations,
intrinsic degeneracies in the neutrino oscillation formulas, the matter density 
uncertainty, and external input on the solar oscillation parameters.\footnote{For a summary of different sources
of errors and their effects, see \Sec~3 of \Ref~\cite{Huber:2002mx}.}
These errors can be reduced by combining different experiments, such as
it has been demonstrated for the intrinsic degeneracies in 
\Refs~\cite{Barger:2002rr,Minakata:2002qi,Burguet-Castell:2002qx,Whisnant:2002fx,Barger:2002xk,Minakata:2002jv}. 
In this spirit we will discuss 
how the synergies of different LBL experiments can be quantified 
as effects which go beyond the increase in integrated luminosity.

Such a discussion makes only sense for 
similar experiments of similar capabilities. In this paper, 
we use two proposed superbeam experiments with equal running times and 
similar levels of sophistication, \ie, the {\sf JHF-SK} and {\sf NuMI} setups. 
It is obvious that the combination of two experiments with similar technology but sizes different by orders
of magnitude will be dominated by the the bigger experiment. It is also obvious that the 
combined fit of more than one experiment will, for comparable setups, be much better than the fit of 
each of the individual experiments. However, the improvement may come from 
better statistics only or from a combination of the better statistics with other 
effects, such as complementary systematics, correlations, and degeneracies. We 
are especially interested in the latter part, since instead of 
building a second experiment, one could also run the first one twice 
as long or built a larger detector.  
The synergy is thus related to the improvement of the measurement 
coming from complementary information of the different experiments. 

It should be clear that we need a method to subtract 
the luminosity increase coming from the combination of different experiments is needed, in
order to compare the results of combined experiments to the ones of the 
individual experiments.
Since we assume equal running times of the individual experiments, we can 
multiply the individual luminosities (running times or target masses) by 
the number of experiments $N$ to be 
combined, and compare the obtained new experiments with the combination of
all experiments operated with their original luminosities.
This method can be easily understood for the two experiments in
this paper: we simulate {\sf JHF-SK} or {\sf NuMI} with twice their original luminosity, \ie,
running time or target mass, and compare it to the combination of the two 
experiments in which the experiments are simulated with their original (single) luminosity. 
We refer to the experiments at double luminosities
further on with {\sf JHF-SK}$_{2L}$ and {\sf NuMI}$_{2L}$.  The difference between the results of
the combined experiments and the results of an individual experiment at double
luminosity then tells us if there is a real synergy effect, which can not only
be explained by the luminosity increase. From the point of view of the
precision of the measurement, it tells us if it were better to run one
experiment twice as long as it was originally intended (or build a twice as big detector) 
or if it were better to build a second experiment instead.\footnote{Of course, there are different arguments to
built two experiments instead of one. However, we focus in this paper on
the physics results only.} To compare, we show the curves 
for {\sf JHF-SK}$_{2L}$ or {\sf NuMI}$_{2L}$ at double luminosity instead of (or in addition to) 
{\sf JHF-SK} and {\sf NuMI} at single luminosity.

\section{The experiments and their simulation}
\label{sec:experiments}

The two experiments considered in this study are the JHF to Super-Kamiokande~\cite{Itow:2001ee} and the
{\sf NuMI}~\cite{Ayres:2002nm} projects using the beams in off-axis configurations, referred to as
``{\sf JHF-SK}'' and ``{\sf NuMI}''. Both 
projects will use the $\rue$ appearance channel in order to measure or improve
 the limits on $\sin^22\theta_{13}$. Neutrino beams produced by meson
decays always contain an irreducible fraction of $\nu_e$, as well as they  have a large high
 energy tail. Therefore, both experiments will exploit the off-axis technology, \ie, building the detector 
slightly off the axis described by the decay pipe, to make the spectrum
much narrower in energy and to suppress the $\nu_e$ component~\cite{offaxis}.
An off-axis beam thus reaches the low background levels necessary for a good sensitivity to the   
$\rue$ appearance signal. Both experiments are planned to be operated at nearly the same $L/E$,
which is optimized for the first maximum of the atmospheric oscillation pattern
for a value of $\dm{31}=3.0\cdot10^{-3}\,\mathrm{eV}^2$.

\begin{table}[ht!]
\begin{center}
\begin{tabular}{|l|c|c|}
\hline
&{\sf JHF-SK}&{\sf NuMI}\\
\hline
\multicolumn{3}{|l|}{Beam} \\
\hline
Baseline&$295\,\mathrm{km}$&$712\,\mathrm{km}$\\
Target Power& $0.77 \, \mathrm{MW}$ & $0.4 \, \mathrm{MW}$ \\
Off-axis angle&$2^\circ$&$0.72^\circ$\\
Mean energy&$0.76\,\mathrm{GeV}$&$2.22\,\mathrm{GeV}$\\
Mean
$L/E$&$385\,\mathrm{km}\,\mathrm{GeV}^{-1}$&$320\,\mathrm{km}\,\mathrm{GeV}^{-1}$\\
\hline
\multicolumn{3}{|l|}{Detector} \\
\hline
Technology&Water Cherenkov&Low-Z calorimeter\\
Fiducial mass&$22.5\,\mathrm{kt}$&$17\,\mathrm{kt}$\\
Running period&5 years&5 years\\
\hline
\end{tabular}
\end{center}
\mycaption{\label{tab:base} The two beams and  detectors as given
in \Refs~{\rm \cite{Itow:2001ee,Ayres:2002nm}} }
\end{table}

Because of the different energies of the two beams, different detector
technologies are used. For the JHF beam, Super-Kamiokande, a water
Cherenkov detector with a fiducial mass of $22.5\,\mathrm{kt}$, is used.
The Super-Kamiokande detector has an excellent electron muon separation
and NC (neutral current) rejection. For the {\sf NuMI} beam, a low-Z calorimeter with a fiducial
mass of $17\,\mathrm{kt}$ is planned, because the hadronic fraction of the
energy deposition is much larger at those energies. In spite of the very
different detector technologies, their performances in terms of background levels and
efficiencies are rather similar. The actual numbers for these quantities and the respective energy
resolution of the detectors can be found in \App. For
a quick comparison, we give in \Tab~\ref{tab:rates} the signal and background
rates for the two experiments at the CHOOZ bound of $\sin^22\theta_{13}=0.1$.
\begin{table}[htb!]
\begin{center}
\begin{tabular}{|l|c|c|}
\hline
&{\sf JHF-SK}&{\sf NuMI}\\
\hline
Signal&137.8&132.0\\
Background&22.6&19.0\\
\hline
\end{tabular}
\end{center}
\mycaption{\label{tab:rates} The signal and background rates for the {\sf JHF-SK} and {\sf NuMI} experiments at
$\sin^22\theta_{13}=0.1$, $\sin^22\theta_{12}=0.8$,  $\sin^22\theta_{23}=1$,
$\dm{31}=3.0\cdot10^{-3}\,\mathrm{eV}^2$,
$\dm{21}=5.0\cdot10^{-5}\,\mathrm{eV}^2$ and $\deltacp=0$ for five years of
neutrino running.
}
\end{table}
The event numbers are similar, but there are subtle differences
in their origin and composition. For example, the fraction of QE (quasi-elastic scattering) 
events is much larger in the {\sf JHF-SK} sample, whereas the matter effect is much more present in the
{\sf NuMI} sample. In fact, the matter effect strongly enhances the {\sf NuMI} signal
compared to the {\sf JHF-SK} signal.

For the calculation of the event rates, we follow the procedure
described in detail in Appendix~A of \Ref~\cite{Huber:2002mx}. Basically we
fold the fluxes as given in \Ref~\cite{Itow:2001ee,fluxes} with the cross
sections from \Ref~\cite{MESSIER}, while we use the energy resolution functions as defined
in \Ref~\cite{Huber:2002mx} in order to obtain realistic detector simulations. 
The probability calculations are performed numerically within a full three flavor scheme, taking into account
the matter effects in a constant average density profile. The parameterizations of
the energy resolution functions and the efficiencies can be found in
\App. 

The evaluation of the precision or the sensitivity for any quantity of
interest requires a statistical analysis of the simulated data because of the low rate counts.
Thus, we employ a $\chi^2$-based method which is  described in
Appendix~C of \Ref~\cite{Huber:2002mx}. We essentially fit all existing information simultaneously, 
which especially means that information from the appearance and disappearance channels
is used at the same time in order to extract as much information as possible.
Beyond the statistical errors, we summarized in \Sec~3 of \Ref~\cite{Huber:2002mx} that there are 
other additional sources of errors limiting the precision of the measurement. First, we take into
account systematical errors (cf. \App), since especially background uncertainties can limit the precision of
superbeam experiments with good statistics. Second, we employ the full multi-parameter correlations among the
 six oscillation parameters plus the matter density, which sizably increase the error bars.
Third, we use external information on the solar parameters and
the matter density. In the appearance probabilities, the solar parameters enter,
up to the second order, only via the product $\dm{21} \cdot \sin2\theta_{12}$.
Therefore, we include for this product an error of $15\%$ which is supposed to come from the KamLAND experiment~\cite{BARGER,Gonzalez-Garcia:2001zy}. 
For the average matter density $\rho =2.8\,\mathrm{g}\,\mathrm{cm}^{-3}$, we assume a precision of $5\%$~\cite{Geller:2001ix}. This
 value contains the contribution of a possible large error on the first Fourier coefficient of
the profile as described in \Refs~\cite{Ota:2000hf,Ota:2002fu} and is shown to be a rather
conservative estimate for more complicated models~\cite{Shan:2001br,Jacobsson:2001zk,Shan:2002px,Jacobsson:2002nb}.
The other oscillation parameters, such as the atmospheric parameters, are measured by the appearance or disappearance rates of the experiments itself such that no external input is needed. Finally, we include in our analysis very carefully  the three degeneracies present in neutrino oscillations (see next section). We essentially follow the methods in \Ref~\cite{Huber:2002mx}, which will be described later in greater detail.

\section{The neutrino oscillation framework}
\label{sec:framework}

Our results are based on a complete numerical analysis including matter effects. It is nevertheless 
useful to have a qualitative understanding of the relevant effects.
We assume standard three neutrino mixing and parameterize the
leptonic mixing matrix $U$ in the same way as the quark mixing matrix~\cite{PDG}. 
In order to obtain intuitively manageable expressions, one can expand the general oscillation
probabilities in powers of the small mass hierarchy parameter $\alpha \equiv
\dm{21}/\dm{31}$ and the small mixing angle $\sin 2\theta_{13}$. Though one can
also include matter effects in such an approximation, we will, for the sake of
simplicity, use here the formulas in vacuum~\cite{Freund:2001ui,FREUND,CERVERA}. 
Note that the expansion in $\sin 2\theta_{13}$ is always good, since this mixing angle is
known to be small. However, the expansion in $\alpha$ is only a good approximation as
long as
the oscillation governed by the solar mass squared difference is small
compared to the leading atmospheric oscillation, \ie,
$\alpha \, \Delta \lesssim 1$. It turns out that the expansion in $\alpha$ can
therefore be used for baselines below 
$L \ll 8000\,\mathrm{km} \left( E_\nu / \GeV \right) \left( 10^{-4}\eV^2 / \dm{21}\right)$,
which is fulfilled for the experiments under consideration if $\Delta m_{21}^2$ is small
enough. The leading terms in small quantities are for the vacuum appearance
probabilities and disappearance
probabilities\footnote{Terms up to the second order, \ie, proportional to
$\sin^2 2\theta_{13}$, $\sin 2\theta_{13}\alpha$, and $\alpha^2$, are taken into
account for $P_{e \mu}$, and terms in the first order are taken into account for
$P_{\mu\mu}$.}
\begin{eqnarray}
\label{eq:PROBVACUUM}
P_{e \mu} & \simeq & \sin^2 2\theta_{13} \, \sin^2 \theta_{23}
\sin^2 {\Delta} \nonumber \\
&\mp&  \alpha\; \sin 2\theta_{13} \, \sin\deltacp  \, \cos\theta_{13} \sin
2\theta_{12} \sin 2\theta_{23}
\sin^3{\Delta} \nonumber \\
&-&  \alpha\; \sin 2\theta_{13}  \, \cos\deltacp \, \cos\theta_{13} \sin
2\theta_{12} \sin 2\theta_{23}
 \cos {\Delta} \sin^2 {\Delta} \nonumber  \\
&+& \alpha^2 \, \cos^2 \theta_{23} \sin^2 2\theta_{12} \sin^2 {\Delta}, \\
\label{eq:DISPROBVACUUM}
P_{\mu \mu} & \simeq & 1 - \cos^2 \theta_{13}
\sin^2 2\theta_{23} \sin^2 {\Delta} \nonumber \\
&+&  2 \alpha  \cos^2 \theta_{13} \cos^2 \theta_{12}
\sin^2 2\theta_{23} {\Delta} \cos{\Delta}.
\end{eqnarray}
The actual numerical values of
$\alpha$ and $\sin^2 2 \theta_{13}$ give each term in
\eqs~(\ref{eq:PROBVACUUM}) and (\ref{eq:DISPROBVACUUM}) a relative weight. In
the LMA case, we have $\alpha\simeq 10^{-2 \pm 1}$. Thus, 
in terms of $\sin^2  2 \theta_{13}$, the first term in \eq~(\ref{eq:PROBVACUUM})
is dominating for $\sin^2  2 \theta_{13} \simeq 0.1$ close to the
CHOOZ bound. For smaller values of $\sin^2  2 \theta_{13}$ all terms are 
contributing equally, until for very small values $\sin 2 \theta_{13} \ll 
\alpha$ the $\alpha^2$ term is dominating.
For long baselines, such as the proposed {\sf NuMI} baselines, the above formulas have 
to be corrected by matter 
effects~\cite{Wolfenstein:1978ue,Wolfenstein:1979ni,Mikheev:1985gs, 
Mikheev:1986wj}. Analytical expressions for the transition probabilities in 
matter can 
be found in \Refs~\cite{FLPR,FREUND,CERVERA}. It turns out that especially the 
first term in \eq~(\ref{eq:PROBVACUUM}) is strongly modified by resonant conversion in matter.

Another important issue for three-flavor neutrino oscillation formulas are
parameter degeneracies, implying that one or more degenerate parameter solutions 
may fit the reference rate vector at the chosen confidence level. In total, there
are three independent two-fold degeneracies, \ie,
an overall ``eight--fold'' degeneracy~\cite{Barger:2001yr}:
\begin{description}
\item[The $\mathrm{sgn}(\Delta m_{31}^2)$ degeneracy:] A degenerate solution for the
opposite sign of $\Delta m_{31}$~\cite{Minakata:2001qm} can often be found close to the $\deltacp$-$\theta_{13}$-plane
(other parameters fixed), but does not necessarily have to lie exactly in the plane. This degeneracy can, in principle, be lifted by matter effects. It is the most important degeneracy in this study.
\item[The $(\deltacp,\theta_{13})$ ambiguity:] 
The $(\deltacp,\theta_{13})$ ambiguity, which allows a degenerate solution in the
$\deltacp$-$\theta_{13}$-plane for the same sign of $\Delta m_{31}^2$, is, in principle, always
present~\cite{Burguet-Castell:2001ez}. However, it depends on the combination of
the oscillation channels if it appears as disconnected solution or not. It is, 
for our study, only relevant as disconnected solution if neutrino and
antineutrino channels are used simultaneously. Otherwise it appears only as one very large solution 
which closely follows the iso-event rate curve in the 
$\deltacp$-$\theta_{13}$-plane.\footnote{For {\sf JHF-SK}, the difference between the original and degenerate solutions
is smaller than 10\% in $\mathrm{log} (\sin^2 2 \theta_{13})$ and for {\sf NuMI}, it is approximately 50\%. Thus, the
degenerate solutions are essentially removed by the combination of the two experiments.}
The position of the second solution can be found almost exactly in the $\deltacp$-$\theta_{13}$-plane 
and is given by the intersection of neutrino and antineutrino iso-event rate curves. It is
therefore relatively easy to find.
\item[The $(\theta_{23},\pi/2-\theta_{23})$ degeneracy:]
For $\theta_{23} \neq \pi/4$, a degenerate solution exists at approximately
$\theta^\prime_{23} = \pi/2-\theta_{23}$. It is not relevant to this work, since
we choose the current atmospheric best-fit value $\theta_{23} = \pi/4$~\cite{Gonzalez-Garcia:2002mu}.
\end{description}
Once degenerate solutions exist, one has to make sure that they are
included in the final result in an appropriate way. We will discuss this issue 
in \Sec~\ref{sec:alternatives} and we will see that especially the $\mathrm{sgn}(\Delta m_{31}^2)$ degeneracy
affects the precision of the results. The role of the degeneracies and the potential to resolve them has, for example,
been studied in \Refs~\cite{Barger:2002rr,Minakata:2002qi,Burguet-Castell:2002qx,Whisnant:2002fx,Barger:2002xk}.

Since we are in this paper mainly interested in the measurements of $\sin^2 2 
\theta_{13}$, the sign of $\Delta m_{31}^2$, and CP violation, we want to 
demonstrate the expected qualitative behavior.
The measurement of $\sin^2 2 \theta_{13}$ is, for small values of $\Delta 
m_{21}^2$ or $\alpha$, dominated by the first term in the appearance probability 
in \eq~(\ref{eq:PROBVACUUM}). This first term is especially influenced by 
matter effects and for our experiments the resonant matter conversion is closest
to the atmospheric mass squared difference. Thus, we expect the {\sf NuMI} experiment to be 
much more affected by $\Delta m_{31}^2$ than the JHF experiment because of 
the longer baseline. In addition, one can see that for larger values of $\Delta 
m_{21}^2$ the second and third terms in \eq~(\ref{eq:PROBVACUUM}) become more 
relative weight. Both terms contain products of $\sin^2 2 \theta_{13}$ and sine 
or cosine of $\deltacp$ and thus it turns out that the correlation with the CP 
phase becomes relevant in this regime. For very large values of $\Delta 
m_{21}^2$, the fourth term becomes dominating and destroys the $\sin^2 2 
\theta_{13}$ sensitivity. However, it is important to note that the 
approximation in the above formulas breaks down for very large $\Delta 
m_{21}^2$. In fact, one can show that the $\sin^2 2 \theta_{13}$ sensitivity 
reappears due to the oscillatory behavior for the investigated experiments 
for $\Delta m_{21}^2 \gtrsim 6 \cdot 10^{-4}  \, \mathrm{eV}^2$. 
We will, however, not discuss this issue in greater detail.

The sensitivity to the sign of $\Delta m_{31}^2$ is mainly spoilt by the 
$\mathrm{sgn} (\Delta m_{31}^2)$ degeneracy. In matter, this degeneracy can 
essentially be resolved by the first term of \eq~(\ref{eq:PROBVACUUM}), because the
matter effects are sizable in this term (if written in terms of quantities in matter).
Thus, one may expect that for small values of $\Delta 
m_{21}^2$ or $\alpha$ and large values of $\sin^2 2 \theta_{13}$ the sensitivity 
to the sign of $\Delta m_{31}^2$ is largest, which is exactly what we find as
qualitative behavior.

The sensitivity to CP violation is dominated by the second and third terms in 
\eq~(\ref{eq:PROBVACUUM}), which require large values of both 
$\alpha$ and $\sin^2 2 \theta_{13}$. However, none of the two parameters should be much 
bigger than the other, since in this case either the first or the fourth 
term becomes too large and spoils the CP violation measurement.
Note that our analysis takes into account the
complete dependence on the CP phase, involving the CP odd $\sin \deltacp$--term
and the CP even $\cos \deltacp$--term.

Summarizing these qualitative considerations, the most important point to note 
is that small values of $\Delta m_{21}^2$ should favor a measurement of the sign 
of $\Delta m_{31}^2$ and large values of  $\Delta m_{21}^2$ a measurement of CP 
violation, while with the initial stage superbeam setups 
used in this paper, a simultaneous sensitivity to the sign of $\Delta m_{31}^2$ 
and CP violation will be hard to achieve.

All results within this work are, unless
otherwise stated, calculated for the current best-fit values
of the atmospheric~\cite{Gonzalez-Garcia:2002mu,Maltoni:2000ib}
and solar neutrino experiments~\cite{Bahcall:2002hv}.
The ranges are, at the $3 \sigma$ confidence level, given by
\begin{eqnarray}
\dm{31} & =& 3_{-2}^{+3}\cdot10^{-3}\,\mathrm{eV}^2\,, \nonumber \\
\sin^2 2\theta_{23} & = & 1_{-0.2}^{+0}\,, \nonumber \\
\dm{21} & = & 5_{-2.7}^{+32}\cdot10^{-5}\,\mathrm{eV}^2\,, \nonumber \\
\sin^2 2\theta_{12} & = & 0.8_{-0.2}^{+0.2}\,, \nonumber
\label{eq:params}
\end{eqnarray}
where the mean values are throughout the text referred to as the ``LMA
solution''. Where relevant, the regions excluded by the indicated ranges
are gray-shaded in our figures. For $\sin^2
2\theta_{13}$ we will only allow values below
the CHOOZ bound~\cite{Apollonio:1999ae}, \ie, $\sin^2 2 \theta_{13} \lesssim 0.1$. 
For the CP phase, we do not make special assumptions, \ie, it can take any
value between zero and $2 \pi$. As indicated above,
this parameter set implies that we do not have to take 
into account the degenerate solution in $(\theta_{23},\pi/2-\theta_{23})$, because this
is only observable for $\sin^2 2 \theta_{23} \neq 1$.

\section{{\sf JHF-SK} and {\sf NuMI} as proposed}
\label{sec:initialstage}

We dicuss now the {\sf JHF-SK} and {\sf NuMI} experiments separately, as well 
as their combination as they are planned. In this section we refer to the 
{\em initial stage} 
experiments as proposed in the Letters of 
Intent~\cite{Itow:2001ee,Ayres:2002nm}, whereas in the following chapters we will study modified setups. Since there are still several options 
for {\sf NuMI} sites, \ie, for baseline and off-axis angle, we here choose, such as in the 
JHF case, the site within the first oscillation maximum at a baseline
of $712 \, \mathrm{km}$ and an off-axis angle of $0.72^\circ$. In addition, we 
choose equal running times of five years of neutrino running only, \ie, we do not 
assume running times with inverted polarities in this section. Under these
conditions, the two experiments should be most comparable.
Built with these parameters, both experiments could measure the atmospheric oscillation parameters 
$\theta_{23}$ and $\Delta m_{31}$ with a good  precision and would be 
sensitive to $\sin^2 2 \theta_{13}$ much below the CHOOZ bound. 

We are most interested in the parameters which are most difficult to 
measure and we therefore first dicuss the $\sin^2 2 \theta_{13}$ sensitivity. It 
is important to note that both experiments with the setups in 
\Refs~\cite{Itow:2001ee,Ayres:2002nm} are optimized for the current atmospheric 
best-fit value $\Delta m_{31}^2 \simeq 3.0 \cdot 10^{-3} \, \mathrm{eV}^2$. 
However, this parameter has still a quite sizable error 
and the $\sin^2 2 \theta_{13}$ sensitivity limit 
strongly depends on its true value determining the 
position of the first oscillation maximum. Before we return to this issue, we 
summarize the results for the current best-fit value $\Delta m_{31}^2 = 3.0 \cdot 10^{-3} \, \mathrm{eV}^2$. 
\begin{figure}[ht!]
\begin{center}
\includegraphics[width=10cm]{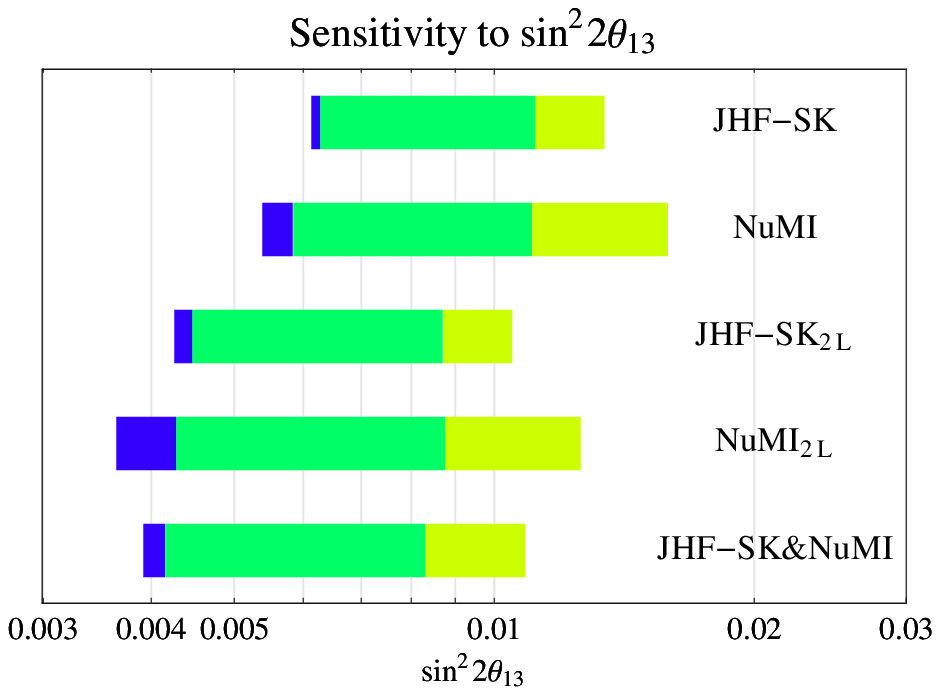}
\end{center}
\mycaption{\label{fig:statusquo} The $\sin^2 2 \theta_{13}$ sensitivity limits
for the {\sf JHF-SK} and {\sf NuMI} experiments at normal and double luminosity (2L), as well as their combination ($90 \%$ confidence level, LMA values). The left edges of the bars correspond to the sensitivity limits from statistics only. The  right edges of the bars correspond to the real sensitivity limits after switching on systematics (blue), correlations (green), and degeneracies (yellow) successively from the left to the right.} 
\end{figure}
Figure~\ref{fig:statusquo} shows the $\sin^2 2 \theta_{13}$ sensitivity limits for the {\sf JHF-SK} and {\sf NuMI} experiments at single and double luminosities and for their combination (with the individual experiments at single luminosity). The left edges of the bars correspond to the sensitivity limits from statistics, while the right edges correspond to the sensitivity limits after successively switching on systematics, correlations, and degeneracies (from left to right). We obtain these plots by finding the largest value of $\sin^2 2 \theta_{13}$ which fits the true $\sin^2 2 \theta_{13}=0$ at the selected confidence level. For the cases of correlations and degeneracies, any solution fitting $\sin^2 2 \theta_{13}=0$ has to be taken into account, since it cannot be distinguished from the best-fit solution. It is important to understand that, without better external information at this time, there is no argument to circumvent this definition and thus no pessimistic choice in that, since a limit is by definition a one-sided interval of all values compatible with a null result. The actual sensitivity limits are therefore the rightmost edges of the bars. Nevertheless, we still find it useful to plot the influence of systematics, correlations, and degeneracies, because these plots demonstrate where the biggest room for optimization is.

For the separate {\sf JHF-SK} and {\sf NuMI} experiments at normal luminosity, \fig~\ref{fig:statusquo} demonstrates that both experiments are approximately equal before the inclusion of systematics, correlations, and degeneracies (left edges of bars). In fact, the {\sf NuMI} experiment performs somewhat better. However, the real sensitivity limits at the right edges of the bars are slightly different, \ie, the {\sf NuMI} experiment is somewhat worse, especially because of the $\mathrm{sgn} (\Delta m_{31}^2)$ degeneracy. In order to discuss the synergy effects of their combination, we argued in \Sec~\ref{sec:synergies} that comparing the experiments at the normal (single) luminosity with their combination does mainly take into account the luminosity increase, \ie, the better statistics. Thus, we show in \fig~\ref{fig:statusquo} the experiments at double luminosity and their combination. It turns out that their combination (at single luminosity) is approximately as good as the {\sf JHF-SK} experiment at double luminosity, mainly due to the reduction of the degeneracy error in the {\sf NuMI} experiment. Nevertheless, this sort of analysis demonstrates that there are no big surprises to expect in building experiments optimized for a similar $L/E$. Obviously, both experiments combined will do better than each of the individual experiments because of the better statistics, but we do not observe a synergy effect when comparing the combination to the individual double luminosity upgrades.

\begin{figure}[ht!]
\begin{center}
\includegraphics[width=10cm]{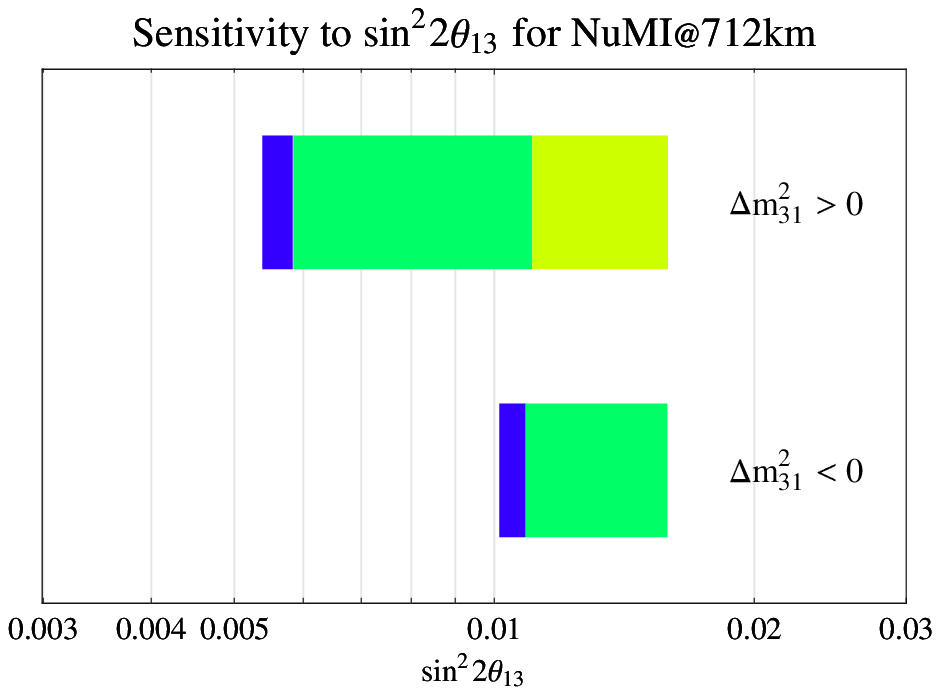}
\end{center}
\mycaption{\label{fig:negsdm} The $\sin^2 2 \theta_{13}$ sensitivity limits
for the {\sf NuMI} experiment for positive and negative signs of $\Delta m_{31}^2$ ($90 \%$ confidence level, LMA values). The left edges of the bars correspond to the sensitivity limits from statistics only. The  right edges of the bars correspond to the real sensitivity limits after switching on systematics (blue), correlations (green), and degeneracies (yellow) successively from the left to the right.} 
\end{figure}
Another interesting issue is that, in this definition, the final sensitivity limits for positive and negative signs of $\Delta m_{31}^2$ are equal because the true rate vectors for $\sin^2 2 \theta_{13}=0$ are identical. This is demonstrated in \fig~\ref{fig:negsdm} for the example of the {\sf NuMI} experiment and can be 
understood by looking at the sensitivity limits 
for the different signs: for the positive sign of $\Delta m_{31}^2$, the purely statistical sensitivity limit (left edge of bars) is somewhat better than the one for the negative sign of $\Delta  m_{31}^2$. After the inclusion of correlations, there is still a degenerate solution with the negative sign of $\Delta m_{31}^2$ making the sensitivity limit worse, which appears as degeneracy part in the bars. From the point of view of a negative sign of $\Delta  m_{31}^2$, however, the degenerate solution appears at $+ |\Delta  m_{31}^2|$ and is somewhat better than the sensitivity limit at the best-fit point at $-|\Delta  m_{31}^2|$ . Thus, there is no degeneracy which makes the solution worse. Both the sensitivity limits for  $\pm |\Delta  m_{31}^2|$ are for our experiments therefore determined by the negative sign solution leading to an equal final sensitivity limit. We will therefore show later only the results for $+|\Delta m_{31}^2|$ and keep in mind that the final sensitivity limits for $\pm |\Delta m_{31}^2|$ are equal.

\begin{figure}[ht!]
\begin{center}
\includegraphics[width=9cm]{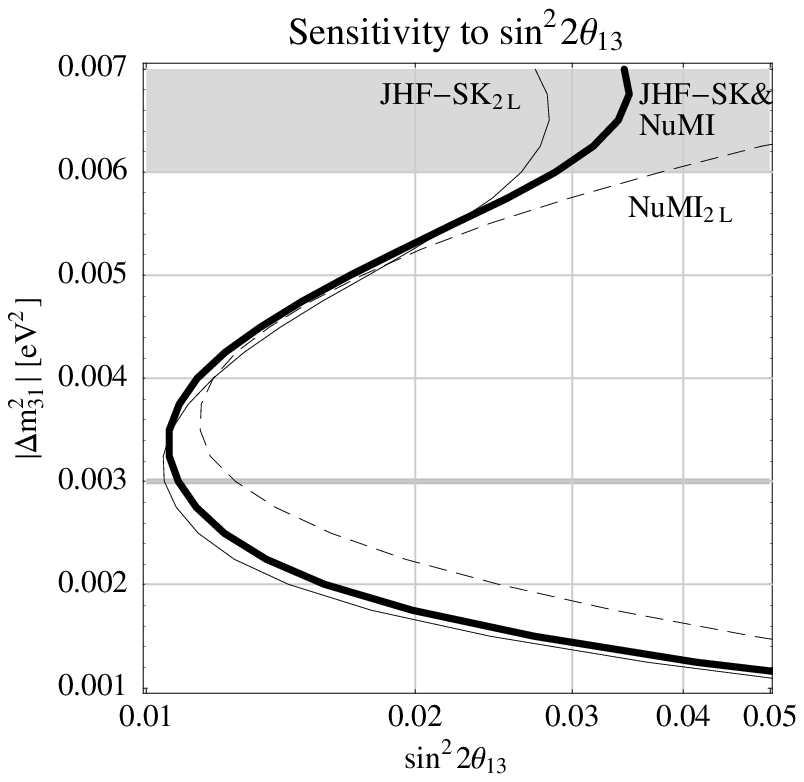}
\end{center}
\mycaption{\label{fig:dm31dep} The $\sin^2 2 \theta_{13}$ excluded regions (on the right-hand sides of the contours) as function of $\Delta m_{31}^2$ (90\% confidence level, LMA values). The curves show {\sf JHF-SK}$_{2L}$ (thin solid curve), {\sf NuMI}$_{2L}$ (thin dashed curve), and {\sf JHF-SK}
and {\sf NuMI} combined (thick curve), where {\sf JHF-SK}$_{2L}$ and {\sf NuMI}$_{2L}$ refer to {\sf JHF-SK} and {\sf NuMI}, respectively, at double luminosity. The atmospheric excluded region is shaded in gray as well as the best-fit value is shown as the gray line. The shown sensitivity limits include systematics, correlations, and degeneracies.} 
\end{figure}

As already mentioned, the $\sin^2 2 \theta_{13}$ sensitivity strongly depends on the true value of $\Delta m_{31}^2$. One may ask, if both experiments combined could reduce the risk of not knowing $\Delta m_{31}^2$ precisely, \ie, flatten the curve describing the $\Delta m_{31}^2$ dependence of the $\sin^2 2 \theta_{13}$ sensitivity limit.
This dependence is shown in \fig~\ref{fig:dm31dep} for the experiments at double luminosity and their combination in order to be immediately able to compare the performance of the individual experiments to their combination. It includes systematics, correlations, and degeneracies.  
First of all, it demonstrates that both experiments are optimized for the current best-fit value $\Delta m_{31}^2 \simeq 3.0 \cdot 10^{-3} \, \mathrm{eV}^2$, since in this region both experiments perform best. At somewhat larger values of $\Delta m_{31}^2$, {\sf NuMI} is slightly better, and at the best-fit value and somewhat smaller values of $\Delta m_{31}^2$, {\sf JHF-SK} is somewhat better, essentially because of the $\mathrm{sgn} (\Delta m_{31}^2)$ degeneracy affecting the {\sf NuMI} results. In addition, it shows that the combination of the two experiments would be almost as good as the best of the two. There is thus no real synergy, \ie, one could  (theoretically) run {\sf JHF-SK} ten years instead of five or use twice the detector mass in order to obtain similar results to the combination of both experiments. 

Two other parameters could be of interest to superbeam experiments: the sign of $\Delta m_{31}^2$ and (maximal) CP violation, as a representative of the $\deltacp$ measurements.
However, for the setups discussed in this section, we find only marginal sensitivities. The sign of $\Delta m_{31}^2$ can not be measured at all with the setups in this section, neither with the individual initial stage {\sf JHF-SK} or {\sf NuMI} experiments, nor with their double luminosity upgrades, nor with their combination. In addition, the CP violation sensitivity is, in the individual experiments, essentially not present in the LMA allowed region. For the combination of the experiments, we only obtain a marginal sensitivity at very high values of $\Delta m_{21}^2$. For a comparison of the {\sf JHF-SK} to other long baseline experiments, such as the JHF to Hyper-Kamiokande upgrade and neutrino factories, see also \Ref~\cite{Huber:2002mx}.

In summary, we have seen that the {\sf JHF-SK} and {\sf NuMI} experiments, as they are planned right now, both have a quite similar $\sin^2 2 \theta_{13}$ sensitivity much below the CHOOZ bound, bot no sensitivity to the sign of $\Delta m_{31}^2$ and only a marginal sensitivity to CP violation. Thus, both experiments are essentially doing the same sort of physics. In the next section, we will thus raise the question if a combination of both experiments under doable modifications can do a much richer physics, \ie, measure the sign of $\Delta m_{31}^2$ or CP violation.

\section{Alternative options for the combination of the experiments}
\label{sec:alternatives}

It is interesting to study if it is possible to do more ambitious physics with the initial stage {\sf NuMI} and {\sf JHF-SK} experiments together than we found in the last section, \ie, if it is possible to measure the sign of $\Delta m_{31}^2$ or CP violation. The price to pay for that could be that (at least) one of the experiments performs worse with respect to the main parameters $\Delta m_{31}^2$, $\theta_{23}$, and $\sin^2 2 \theta_{13}$. Thus, we will first discuss alternative options for both experiments in order to measure the sign of $\Delta m_{31}^2$ or CP violation. Then, we will investigate what modifications of the experiments mean for the sensitivity limit of $\sin^2 2 \theta_{13}$ as one of the
most important goals of these experiments.

Before we come to the individual parameter measurements, we need to clarify what we understand under {\em alternative options} for the initial stage experiments. Since the Super-Kamiokande detector exists already and the {\sf NuMI} decay pipe is already built, the most reasonable options for modifications are the {\sf NuMI} baseline (with a certain off-axis angle) and the possibility of antineutrino running for both experiments, where the {\sf NuMI} baseline length and off-axis angle cannot be optimized independently because of the already installed decay pipe. For a fixed off-axis angle, the possible detector sites lie on an ellipse on the Earth's surface. Since, besides other constraints, a possible detector site requires infrastructure and it turns out that the physics potential of both experiments is essentially in favor of a longer {\sf NuMI} baseline, we choose two additional candidates with a baseline much longer than $712 \, \mathrm{km}$. One is at a baseline length of $L=890 \, \mathrm{km}$ at an off-axis angle of $0.72^\circ$, which is the longest possible choice for this off-axis angle. 
\begin{table}[ht!]
\begin{center}
\begin{tabular}{|r|cc|cccc|c|}
\hline
 & \multicolumn{2}{c|}{{\sf JHF-SK}}  & \multicolumn{4}{c|}{{\sf NuMI}} & Combined \\
No. & $\nu$ & $\bar{\nu}$ & $\nu$ & $\bar{\nu}$ & L [km] & OA angle & Label\\
\hline
\hline
1 & 1 & 0 & 1 & 0 & 712 & 0.72$^\circ$ & 712$\nu\nu$\\
2 & 1 & 0 & 1 & 0 & 890 & 0.72$^\circ$ & 890$\nu\nu$\\
3 & 1 & 0 & 1 & 0 & 950 & 0.97$^\circ$ & 950$\nu\nu$\\
\hline
4 & 2/8 & 6/8 & 2/7 & 5/7 & 712 & 0.72$^\circ$ & 712cc\\
5 & 2/8 & 6/8 & 2/7 & 5/7 & 890 & 0.72$^\circ$ & 890cc \\
6 & 2/8 & 6/8 & 2/7 & 5/7 & 950 & 0.97$^\circ$ & 950cc \\
\hline
7 & 1 & 0 & 0 & 1 & 712 & 0.72$^\circ$ & 712$\nu\bar{\nu}$\\
8 & 1 & 0 & 0 & 1 & 890 & 0.72$^\circ$ & 890$\nu\bar{\nu}$\\
9 & 1 & 0 & 0 & 1 & 950 & 0.97$^\circ$ & 950$\nu\bar{\nu}$\\
\hline
10 & 0 & 1 & 1 & 0 & 712 & 0.72$^\circ$ & 712$\bar{\nu}\nu$\\
11 & 0 & 1 & 1 & 0 & 890 & 0.72$^\circ$ & 890$\bar{\nu}\nu$\\
12 & 0 & 1 & 1 & 0 & 950 & 0.97$^\circ$ & 950$\bar{\nu}\nu$\\
\hline
\end{tabular}
\end{center}
\mycaption{\label{tab:scenarios}The tested alternative scenarios for the combination of
the {\sf JHF-SK} and {\sf NuMI} experiments. The columns refer to: the scenario number, the neutrino and antineutrino running fractions $\nu$ and $\bar{\nu}$ for the individual experiments, the {\sf NuMI} baseline length $L$, the {\sf NuMI} off-axis angle, and the labels used in the following plots.}
\end{table}
A possible detector site is Fort Frances, Ontario, with a baseline $L=875 \, \mathrm{km}$~\cite{HARRIS}. The other candidate is at a baseline length of $L=950 \, \mathrm{km}$ at an off-axis angle of $0.97^\circ$. It has a longer baseline with a larger off-axis angle, \ie, with larger matter effects and the spectrum is sharper peaked but the statistics is worse. A possible site at exactly this baseline is Vermilion Bay, Ontario. For the running with inverted polarities, we investigate options with neutrino running in both {\sf JHF-SK} and {\sf NuMI}, neutrino running in {\sf JHF-SK} and antineutrino running in {\sf NuMI} or vice versa, and combinations of neutrino and antineutrino running in both {\sf JHF-SK} and {\sf NuMI} with a splitting of the running time such that we approximately have the same numbers of neutrino and antineutrino events, \ie, 2/8 of neutrino running and 6/8 of antineutrino running at {\sf JHF-SK} and 2/7 of neutrino running and 5/7 of antineutrino running at {\sf NuMI}. The tested scenarios are listed in \Tab~\ref{tab:scenarios} and we will further on use the labels shown in the column ``Label'' to identify the different scenarios. The labels refer to the {\sf NuMI} baseline length, the {\sf JHF-SK} polarity (first letter) and the {\sf NuMI} polarity (second letter), where the abbreviation ``$\nu$'' stands for neutrino running only, ``$\bar{\nu}$'' for antineutrino running only, and ``c'' for the combined option.

\subsection{The measurement of the sign of $\Delta m_{31}^2$}

One possibility to broaden the physics potential is to measure the sign of $\Delta m_{31}^2$. Recent studies, such as \Ref~\cite{Huber:2002mx}, demonstrate that it is very hard to access this parameter even at high-luminosity superbeam upgrades (such as JHF to Hyper-Kamiokande) or neutrino factories essentially due to the $\mathrm{sgn}( \Delta m_{31}^2 )$ degeneracy. However, the combination of two complementary baselines can help to resolve this degeneracy. Especially, a very long baseline with large matter effects adds a lot to the sensitivity, though it is not optimal as a standalone setup because of the lower statistics.

Before we present our analysis, we define the sensitivity to the sign of $\Delta m_{31}^2$ and describe how we evaluate it. Sensitivity to a certain sign of $\Delta m_{31}^2$ exists if there is no solution with the opposite sign which fits the true values at the chosen confidence level. It is essential to understand that if we find such a solution, we will not be able to measure the sign of $\Delta m_{31}^2$, because the new solution opens the possibility of a completely new parameter set at $\simeq \, -\Delta m_{31}^2$ which cannot be distinguished from the best-fit one. Thus, finding such a degenerate solution proofs that there is no sensitivity to the sign of $\Delta m_{31}$ at the chosen best-fit parameter values. Practically, we are scanning the half of the  parameter space with the inverted $\Delta m_{31}^2$ for the $\mathrm{sgn}( \Delta m_{31}^2 )$ degeneracy. We do not find the degenerate solution exactly in the $(\deltacp,\theta_{13})$-plane at $-\Delta m_{31}^2$,  since it can lie slightly off this plane. This makes it necessary to use a local minimum finder in the high dimensional parameter space, which, for example, can be started in the $(\deltacp,\theta_{13})$-plane at $-\Delta m_{31}^2$ and then runs down into the local minimum off the plane. In other words, the sign of $\Delta m_{31}^2$ measurement is also correlated with the other parameter measurements. Moreover, it turns out that the actual value of $\deltacp$ strongly determines the sign of $\Delta m_{31}^2$ sensitivity. Since $\deltacp$ will not be measured before the considered experiments, the only information on $\deltacp$ can come from the experiments themselves, \ie, we assume that we do not know $\deltacp$ {\em a priori}. For a certain parameter set, the actual value of $\deltacp$ will determine the confidence level at which the degenerate solution appears. Since, in some sense, without {\em a priori} measurement all values of $\deltacp$ are equally favored, we compute the sign of $\Delta m_{31}^2$ sensitivity for all possible value of $\deltacp$ and take the worst case. Thus, our regions show where we are sensitive to the sign of $\Delta m_{31}^2$ in either case of $\deltacp$. Another issue is that there is a slight difference in the measurement of the sign of $\Delta m_{31}^2$  and the sensitivity to a positive or negative sign of $\Delta m_{31}^2$. We ask for the sensitivity to a positive or negative sign of $\Delta m_{31}^2$ and show the regions were this sign could be verified. To measure the sign of $\Delta m_{31}^2$ would mean to proof its value in either case, \ie, one had to take the worst case  of the sensitivity regions for positive and negative signs.

As a first result of this analysis, we find, as indicated in the last section, no sensitivity to the sign of $\Delta m_{31}^2$ neither for {\sf JHF-SK} or {\sf NuMI} (at $712 \, \mathrm{km}$ with neutrino running only) at single or double luminosity, nor for their combination. However, we find that increasing the {\sf NuMI} baseline leads to sensitivity at baselines of $L \gtrsim 800 \, \mathrm{km}$. The optimum at approximately the second oscillation maximum at $L \sim 1400 \, \mathrm{km}$, which can, however, not be reached by the fixed decay pipe and does not give a good sensitivity to $\sin^2 2 \theta_{13}$ anymore. Thus, we take the above introduced baseline-off-axis angle combinations at $890 \, \mathrm{km}$ and $950 \, \mathrm{km}$ with the neutrino-antineutrino running time combinations of {\sf JHF-SK} and {\sf NuMI} listed in \Tab~\ref{tab:scenarios}.
\begin{figure}[ht!]
\begin{center}
\includegraphics[width=16cm]{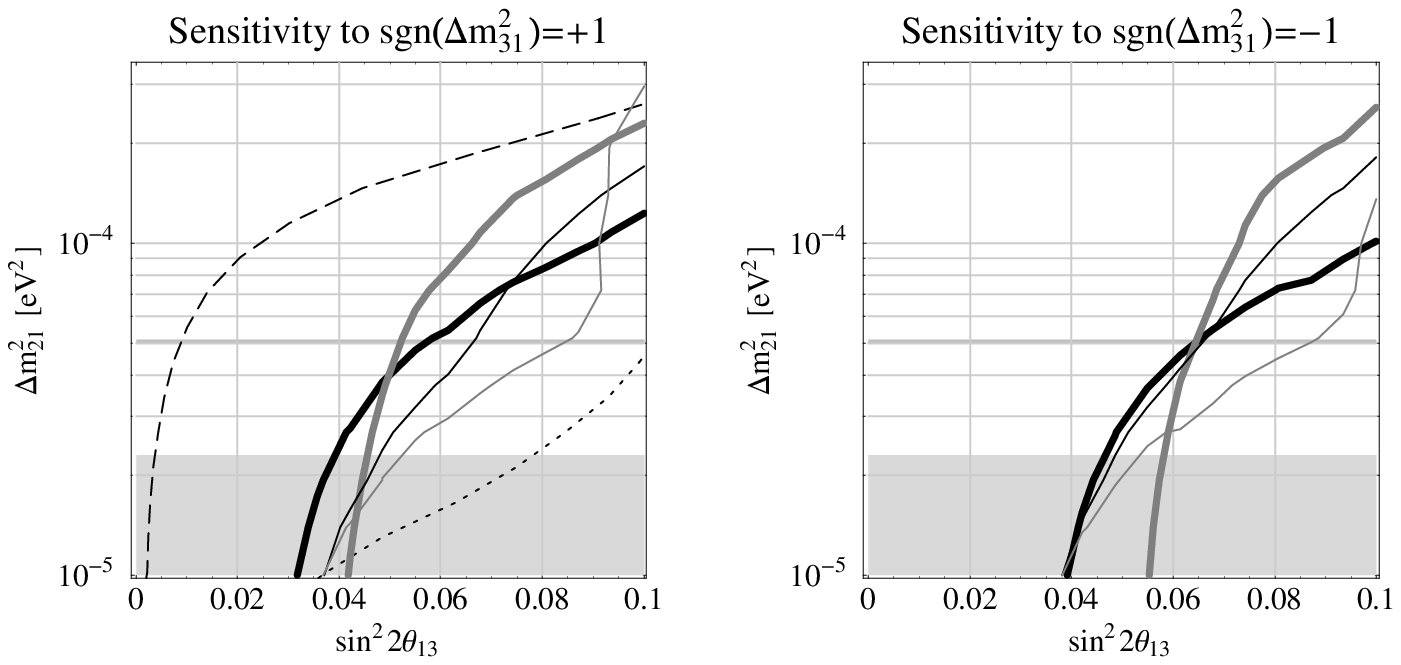}
\end{center}
\mycaption{\label{fig:sgndm} The sensitivity regions to a positive (left-hand plot) and negative (right-hand plot) sign of $\Delta m_{31}^2$ 
 as functions of $\sin^2 2 \theta_{13}$ and $\Delta m_{21}^2$ ($90 \%$ confidence level, LMA values), where sensitivity exists on the right-hand side of the curves. In the plots, the curves for {\sf JHF-SK} and {\sf NuMI} combined are shown for the following options, listed in \Tab~\ref{tab:scenarios}: $890\nu\nu$ (thick gray curve), $950\nu\nu$ (thick black curve), $890cc$ (thin gray curve), and $950cc$ (thin black curve). In addition the NuFact-I (initial stage neutrino factory, dashed curve) and JHF-HK (JHF to Hyper-Kamiokande superbeam upgrade, dotted curve) setups from \Ref~{\rm \cite{Huber:2002mx}} are shown for comparison in the left-hand plot. The gray horizontal line marks the LMA best-fit value, whereas the gray-shaded region corresponds to the current LMA excluded region.}
\end{figure}
Figure~\ref{fig:sgndm} shows the sensitivity to a positive (left-hand plot) or 
negative (right-hand plot) sign of $\Delta m_{31}^2$ for the best of these 
options. We do not show the scenarios with running one of the experiments with 
antineutrinos only, because this reduces the reach in $\Delta m_{21}^2$ 
severely. The reduction of the sensitivity at larger values of $\Delta m_{21}^2$ 
comes, as described in \Sec~\ref{sec:framework}, essentially from the larger 
CP-effects, above a certain
threshold value in $\Delta m_{21}^2$ we are not sensitive to the 
sign of $\Delta m_{31}^2$ anymore.  As the main result, 
we find that running both beams with neutrinos only at a {\sf NuMI} baseline of $890 \, \mathrm{km}$ or
$950 \, \mathrm{km}$ gives the best results, where the sensitivity for a 
negative sign of $\Delta m_{31}^2$ is slightly worse. Above $\Delta m_{21}^2
\simeq 4.5 \cdot 10^{-5} \, \mathrm{eV^2}$ the $890 \, \mathrm{km}$ option with
the smaller
off-axis angle performs better, whereas below $\Delta m_{21}^2 \simeq 4.5 \cdot
10^{-5} \, \mathrm{eV^2}$ the $950 \, \mathrm{km}$ option with the larger 
off-axis angle is better.
\begin{table}[ht!]
\begin{center}
\begin{tabular}{|r|ccc|cccc|cc|}
\hline
 & \multicolumn{3}{c|}{{\sf JHF-SK}}  & \multicolumn{4}{c|}{{\sf NuMI}} & \multicolumn{2}{c|}{Combined} \\
No. & $\nu$ & $\bar{\nu}$ & Sensitivity & $\nu$ & $\bar{\nu}$ & L [km] & Sensitivity & Label & Sensitivity \\
\hline
\hline
1 & 1 & 0 & None & 1 & 0 & 712 & None & 712$\nu\nu$ & None \\
2 & 1 & 0 & None & 1 & 0 & 890 & None & 890$\nu\nu$ & Good \\
3 & 1 & 0 & None & 1 & 0 & 950 & None & 950$\nu\nu$ & Good \\
\hline
4 & 2/8 & 6/8 & None & 2/7 & 5/7 & 712 & Marginal & 712cc & Marginal \\
5 & 2/8 & 6/8 & None & 2/7 & 5/7 & 890 & Marginal & 890cc & Poor \\
6 & 2/8 & 6/8 & None & 2/7 & 5/7 & 950 & Marginal & 950cc & Good \\
\hline
7 & 1 & 0 & None & 0 & 1 & 712 & None & 712$\nu\bar{\nu}$ &Marginal \\
8 & 1 & 0 & None & 0 & 1 & 890 & None & 890$\nu\bar{\nu}$ & Poor \\
9 & 1 & 0 & None & 0 & 1 & 950 & None & 950$\nu\bar{\nu}$ & Poor \\
\hline
10 & 0 & 1 & None & 1 & 0 & 712 & None & 712$\bar{\nu}\nu$ & Marginal \\
11 & 0 & 1 & None & 1 & 0 & 890 & None & 890$\bar{\nu}\nu$ & Poor \\
12 & 0 & 1 & None & 1 & 0 & 950 & None & 950$\bar{\nu}\nu$ & Poor \\
\hline
\end{tabular}
\end{center}
\mycaption{\label{tab:sgnres}The sign of $\Delta m_{31}^2$ sensitivity (qualitatively) for the tested alternative options for the individual {\sf JHF-SK} and {\sf NuMI} experiments (normal luminosity), as well as their combination. The columns refer to: the scenario number, the neutrino and antineutrino running fractions $\nu$ and $\bar{\nu}$ for the individual experiments, the {\sf NuMI} baseline length $L$, and the label of the combined experiments. The sensitivity reaches are classified as ``None'' (no sensitivity), ``Marginal'' (sensitivity only in lower right corner in plots in \fig~\ref{fig:sgndm}), ``Poor'' (good maximal reach in $\sin^2 2 \theta_{13}$ or $\Delta m_{21}^2$, but area covered in plots in \fig~\ref{fig:sgndm} relatively small), and ``Good'' (good coverage in $\sin^2 2 \theta_{13}$ and $\Delta m_{21}^2$ reach as well as area). The best options are also shown in \fig~\ref{fig:sgndm}.}
\end{table}
\Tab~\ref{tab:sgnres} summarizes our performance tests for the sign of $\Delta m_{31}$ qualitatively.
Comparing the sensitivity regions to future high luminosity superbeam upgrades or initial stage neutrino factories (\cf, left-hand plot), we could achieve relatively good sensitivities to the sign of $\Delta m_{31}^2$ even with the initial stage setups. As far as the dependence on the true value of $\Delta m_{31}^2$ is concerned, a larger value of $\dm{31}$ increases the resonance energy and therefore reduces the matter effects. However, a larger value of $\dm{31}$
allows a higher energy in order to overcompensate the smaller matter effects and to finally obtain somewhat better results. The same arguments also work for smaller values of $\dm{31}$. In this case, however, an experiment, even if it is fully optimized for this lower value, strongly suffers from the loss in statistics
since increasing the baseline or lowering the energy strongly reduces the number of events. 

\subsection{The sensitivity to CP violation}

Another alternative for optimization with the combined {\sf JHF-SK} and {\sf NuMI} experiments under modified conditions is leptonic CP violation. We restrict the discussion to maximal CP violation, \ie, $\deltacp= \pm \pi/2$, but one could also discuss connected topics, such as the precision of the $\deltacp$ measurement or the establishment of CP violation for CP phases closer to CP conservation, \ie, $\deltacp= 0$ or $\pi$. However, we expect the qualitative behavior for these problems to be similar and choose the sensitivity to maximal CP violation as representative.

For our analysis, we compute the degenerate solution in $\mathrm{sgn}(\Delta m_{31}^2)$ as discussed in the last section, since it turns out that this degeneracy can have a CP phase very different from the original solution~\cite{Huber:2002mx}. The question of the sensitivity to maximal CP violation can be answered by setting the CP phase to $\deltacp = +\pi/2$ or $\deltacp = -\pi/2$ and computing the four $\chi^2$-values at the best-fit and degenerate solution for $\deltacp = 0$ and $\pi$, respectively. For the case of antineutrino running, the additional possible ambiguity in the $(\deltacp,\theta_{13})$-plane is taken into account.\footnote{However, it does not play an important role here since it does not map $\deltacp=\pm \pi/2$ to zero or $\pi$.} If one of the computed $\chi^2$-values is under the threshold for the selected confidence level, maximal CP violation cannot be established for the evaluated parameter set.

\begin{figure}[ht!]
\begin{center}
\includegraphics[width=9cm]{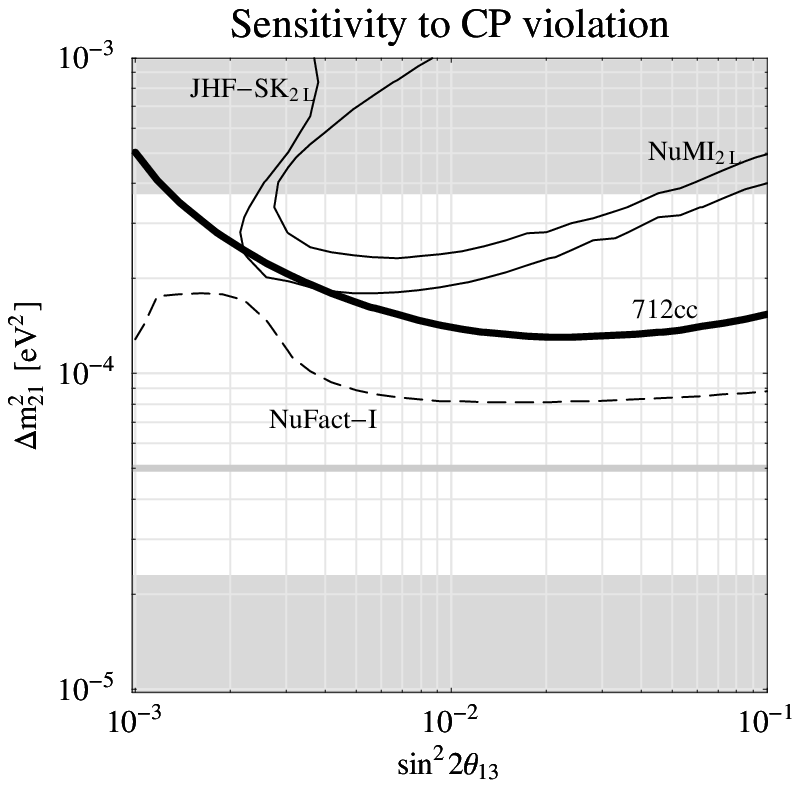}
\end{center}
\mycaption{\label{fig:CPVcomparison} The sensitivity to maximal CP violation $\deltacp=+\pi/2$ for $\Delta m_{31}^2>0$ as function of $\sin^2 2 \theta_{13}$ and $\Delta m_{21}^2$ (90\% confidence level, LMA values). The curves for {\sf JHF-SK}$_{2L}$ and {\sf NuMI}$_{2L}$
at double luminosity with neutrino running only, as well as the curve for both combined ($712cc$) are shown. All curves are plotted for the {\sf NuMI} baseline of $712 \, \mathrm{km}$. In addition, we show the result for the NuFact-I scenario (initial stage neutrino factory) from \Ref~{\rm \cite{Huber:2002mx}} for comparison as dashed curve. The LMA excluded region is shaded in  gray and its best-fit value is marked by the gray horizontal line.}
\end{figure}

\begin{figure}[ht!]
\begin{center}
\includegraphics[width=16cm]{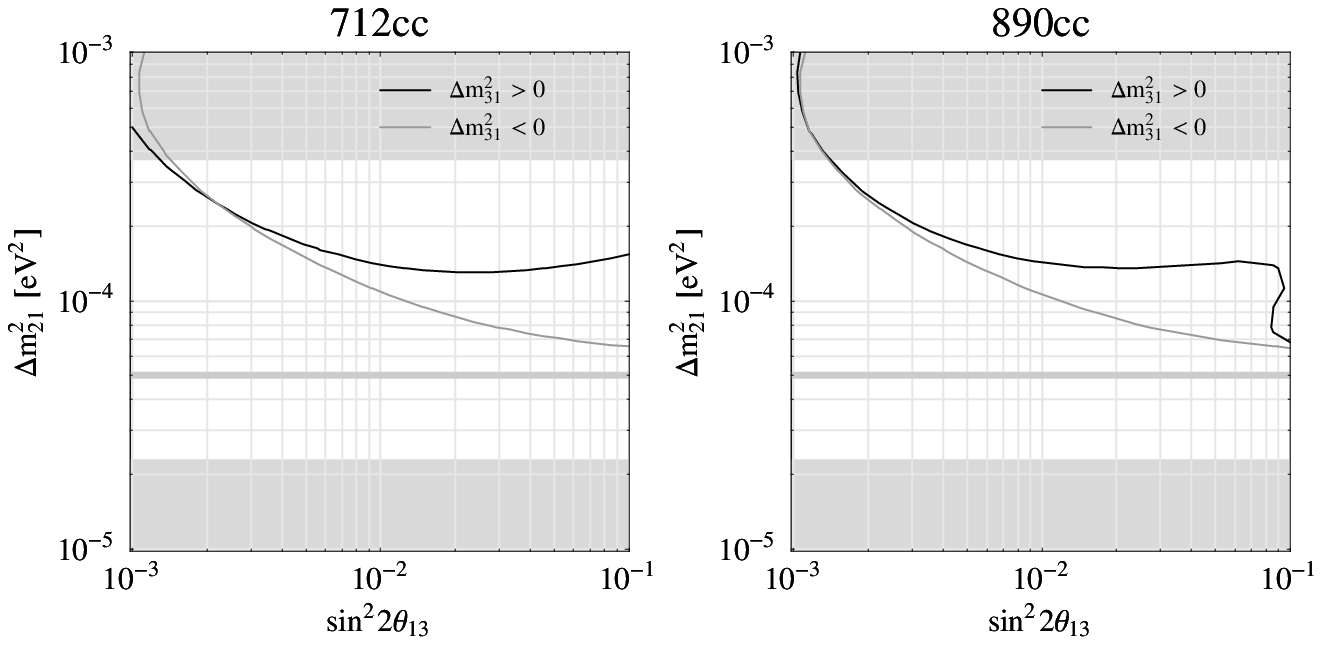}
\end{center}
\mycaption{\label{fig:CPVsens} The sensitivity to maximal CP violation $\deltacp=+\pi/2$ for $\Delta m_{31}^2>0$ (black curves) and $\Delta m_{31}^2<0$ (gray curves) as functions of $\sin^2 2 \theta_{13}$ and $\Delta m_{21}^2$ (90\% confidence level, LMA values). The left-hand plot shows the sensitivity for the combined experiments and {\sf NuMI} at $712 \, \mathrm{km}$ ($712cc$) and the right-hand plot for {\sf NuMI} at $890 \, \mathrm{km}$ ($890cc$), both with combined neutrino and antineutrino running. The LMA excluded region is shaded in gray and its best-fit value is marked by the gray horizontal line.}
\end{figure}

Before we discuss the options introduced in \Tab~\ref{tab:scenarios}, we want to compare the performance of the individual {\sf JHF-SK} and {\sf NuMI} experiments at double luminosity to their combination. \fig~\ref{fig:CPVcomparison} shows the sensitivity to maximal CP violation $\deltacp=+\pi/2$ for $\Delta m_{31}^2>0$ as function of $\sin^2 2 \theta_{13}$ and $\Delta m_{21}^2$, since we find that there are no interesting qualitative differences for $\deltacp=-\pi/2$ or $\Delta m_{31}^2<0$. In this figure, the curves for {\sf JHF-SK}$_{2L}$ and {\sf NuMI}$_{2L}$ (at double luminosity) for neutrino running only, as well as their combination with {\sf NuMI} at $L=712 \, \mathrm{km}$ with combined neutrino-antineutrino running (Scenario ``712cc'' in \Tab~\ref{tab:scenarios}) are shown.
One can easily see that the combination of the two experiments under somewhat modified conditions (antineutrino running included) leads to a much better coverage of the $\sin^2 2 \theta_{13}$-$\Delta m_{21}^2$-plane, which cannot be explained by the statistics increase only. In this case, one could talk about a synergy between the two experiments. It turns out that most of the alternative options including antineutrino running, if not put exclusively to {\sf JHF-SK}, are performing similarly well. Thus, we show in \fig~\ref{fig:CPVsens} the sensitivity to maximal CP violation for the two options of combined neutrino and antineutrino running with a {\sf NuMI} baseline of $712 \, \mathrm{km}$ and $890 \, \mathrm{km}$. 
In this figure, essentially all curves for the scenarios classified as ``Good'' in \Tab~\ref{tab:CPres} summarizing the results of this analysis qualitatively, are very close to the ones plotted for $\Delta m_{31}^2>0$ and $\Delta m_{31}^2<0$, respectively. In addition, there is no big difference between choosing $\deltacp=+\pi/2$ and $\deltacp=-\pi/2$. However, the results for the $950 \, \mathrm{km}$ {\sf NuMI} baseline are slightly worse. In summary, we find that for the detection of maximal CP violation all combinations which include a substantial fraction of antineutrino running perform very well -- except from the option of running {\sf NuMI} with neutrinos only and {\sf JHF-SK} with antineutrinos only.

\begin{table}[ht!]
\begin{center}
\begin{tabular}{|r|ccc|cccc|cc|}
\hline
 & \multicolumn{3}{c|}{{\sf JHF-SK}}  & \multicolumn{4}{c|}{{\sf NuMI}} & \multicolumn{2}{c|}{Combined} \\
No. & $\nu$ & $\bar{\nu}$ & Sensitivity & $\nu$ & $\bar{\nu}$ & L [km] & Sensitivity & Label & Sensitivity \\
\hline
\hline
1 & 1 & 0 & None & 1 & 0 & 712 & None & 712$\nu\nu$ & Poor \\
2 & 1 & 0 & None & 1 & 0 & 890 & None & 890$\nu\nu$ & Poor \\
3 & 1 & 0 & None & 1 & 0 & 950 & None & 950$\nu\nu$ & Poor \\
\hline
4 & 2/8 & 6/8 & None & 2/7 & 5/7 & 712 & None & 712cc & Good \\
5 & 2/8 & 6/8 & None & 2/7 & 5/7 & 890 & None & 890cc & Good \\
6 & 2/8 & 6/8 & None & 2/7 & 5/7 & 950 & None & 950cc & Suboptimal \\
\hline
7 & 1 & 0 & None & 0 & 1 & 712 & None & 712$\nu\bar{\nu}$ & Good \\
8 & 1 & 0 & None & 0 & 1 & 890 & None & 890$\nu\bar{\nu}$ & Good \\
9 & 1 & 0 & None & 0 & 1 & 950 & None & 950$\nu\bar{\nu}$ & Suboptimal \\
\hline
10 & 0 & 1 & None & 1 & 0 & 712 & None & 712$\bar{\nu}\nu$ & Poor \\
11 & 0 & 1 & None & 1 & 0 & 890 & None & 890$\bar{\nu}\nu$ & Poor \\
12 & 0 & 1 & None & 1 & 0 & 950 & None & 950$\bar{\nu}\nu$ & Poor \\
\hline
\end{tabular}
\end{center}
\mycaption{\label{tab:CPres}The CP violation sensitivity (qualitatively) for the tested alternative options for the individual {\sf JHF-SK} and {\sf NuMI} experiments (normal luminosity), as well as their combination. The columns refer to: the scenario number, the neutrino and antineutrino running fractions $\nu$ and $\bar{\nu}$ for the individual experiments, the {\sf NuMI} baseline length $L$, and the label of the combined experiments. The sensitivity reaches are classified in ``None'' (no or only marginal sensitivity in the LMA range), ``Poor'' (good reach in $\sin^2 2 \theta_{13}$ or $\Delta m_{21}^2$, but area covered in plots in \fig~\ref{fig:CPVsens} relatively small), ``Suboptimal'' (fairly good, but not optimal coverage in area), and ``Good'' (good coverage in area). The options classified with ``Good'' correspond to the curves shown in \fig~\ref{fig:CPVsens}.}
\end{table}

The qualitative behavior can be understood from \eq~(\ref{eq:PROBVACUUM}), where the second and third terms contribute most to the CP violation sensitivity. They are both getting the largest relative weight if both $\sin^2 2 \theta_{13}$ and $\Delta m_{21}^2$ are rather large, whereas they are suffering from the large absolute background from the first term if $\Delta m_{21}^2$ is too small. Thus, we could only measure CP violation in the so-called HLMA-region, \ie, at the upper end of the LMA range. As far as the dependence on the value of $\dm{31}$ is concerned, a larger value of $\dm{31}$ would imply a lower value of $\alpha$ in \eq~(\ref{eq:PROBVACUUM}) and thus decrease the relative strength of CP effects. However, it would allow to choose a shorter baseline or higher energy, which both increase the total event rates. This improvement in statistics has more relative weight than the reduction of the CP effects. For lower values of $\dm{31}$, it is very hard to improve the results by optimizing the energy or baseline, since, such as for the sign of $\Delta m_{31}^2$ measurement, the statistics always decreases and thus makes the measurement difficult.

\subsection{The sensitivity to $\sin^2 2 \theta_{13}$}

\begin{figure}[ht!]
\begin{center}
\includegraphics[width=10cm]{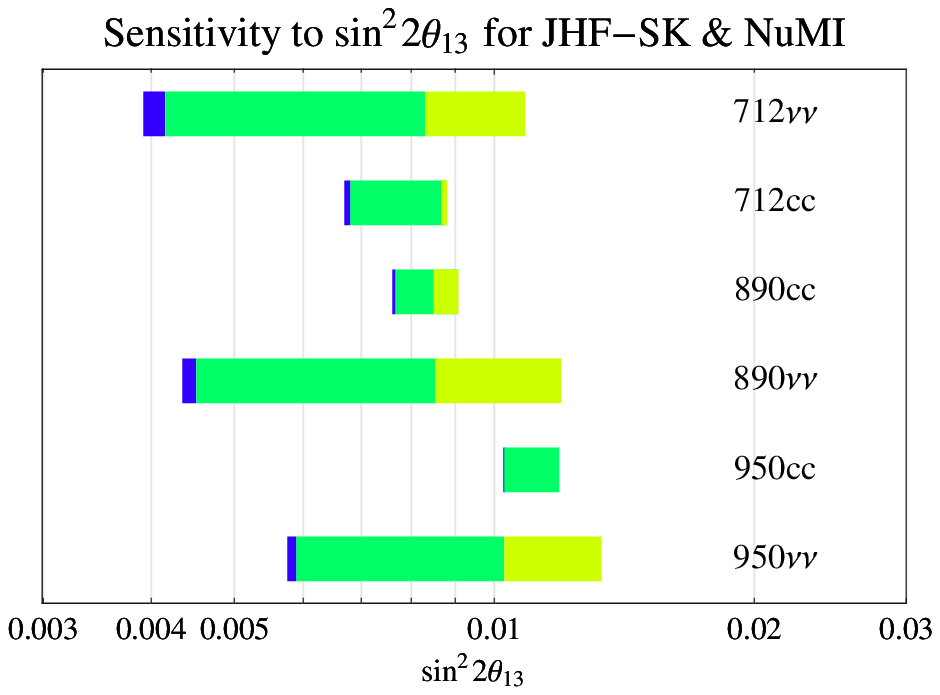}
\end{center}
\mycaption{\label{fig:diffsites} The $\sin^2 2 \theta_{13}$ sensitivity limits ($90 \%$ confidence level, LMA values)
for the combined {\sf JHF-SK} and {\sf NuMI} experiments for a selected set of the alternative options and the original combination (first bar). The labels are defined in \Tab~\ref{tab:scenarios}. The left edges of the bars correspond to the sensitivity limits from statistics only. The  right edges of the bars correspond to the real sensitivity limits after switching on systematics (blue), correlations (green), and degeneracies (yellow) successively from the left to the right.} 
\end{figure}

We have seen that there are several alternative options for the combination of the {\sf JHF-SK} and {\sf NuMI} experiments. We have also indicated that a different setup could make the measurement of $\sin^2 2 \theta_{13}$ worse. Thus, we show in \fig~\ref{fig:diffsites} how much the sensitivity is changed for a selected set of the alternative options as well as the original combination at $712 \, \mathrm{km}$ (first bar).
The sensitivity limit is, in all cases, reduced due to statistics only (left edges of the bars). However, due to the reduction of correlations and degeneracies, \ie, partial complementarity of the different baselines and polarities, many of the setups perform even better after the inclusion of all error sources. In the end, any of the shown setups has a very good performance in the $\sin^2 2 \theta_{13}$ sensitivity. In addition, one has to keep in mind that the dependency on the atmospheric and solar mass squared differences can affect this result much stronger than choosing the suboptimal solution. For example, for larger values of $\Delta m_{21}^2$ these limits are shifted to higher values and for smaller values of $\Delta m_{21}^2$ the limits are shifted to smaller values. At about $\Delta m_{21}^2 \simeq 3 \cdot 10^{-4} \, \mathrm{eV}^2$ we would finally reach the CHOOZ bound. Since this value is still within the LMA allowed region, there is in the worst case no guarantee that a limit better than the CHOOZ could be obtained.

For the alternative options of the combined two experiments, one can again raise the question if it is 
possible to reduce the risk of having optimized for the wrong value of $|\dm{31}|$.  We find that spreading the $L/E$ values of the two experiments, \ie, having one experiment somewhat below the first maximum at $|\dm{31}|=3.0\cdot 10^{-3} \, \mathrm{eV}^2$ and the other above, would only improve the $\sin^2 2 \theta_{13}$ sensitivity limit for $|\dm{31}|>3.0\cdot 10^{-3} \, \mathrm{eV}^2$. This can, in general, be understood in terms of the scaling of the statistics. Increasing the value of $|\dm{31}|$ would mean shifting the first maximum to higher energies or shorter baselines. For the case of higher energies, the
event rates, scaling $\propto E^{3+x}$, would be increased.\footnote{Here the factor $E^2$ comes from the flux of an
off-axis beam and an additional factor $E^{1+x}$ comes from the energy dependence of the
cross section.} For the case of shorter baselines, the rates simply scale $\propto1/L^2$.
Thus, any of the two options would lead to higher event rates for an
experiment optimized for a larger value of $|\dm{31}|$. The same reasoning for lower values of $|\dm{31}|$ 
always results in much lower event
rates. Thus, even an experiment optimized for this lower value of $|\dm{31}|$ tends to perform much
worse in absolute numbers. From calculations, we find that for $|\dm{31}|=5.0\cdot
10^{-3} \, \mathrm{eV}^2$ the sensitivity limit could be relatively improved by $\simeq 50\%$ (linear scale) 
compared to the standard setup 712$\nu\nu$ without loosing much at lower
values of $|\dm{31}|$. This result is, however, obtained for unrealistic values
of the baselines, which means that for a realistic setup the improvement will be smaller.  For  
$|\dm{31}|=1.0\cdot 10^{-3} \, \mathrm{eV}^2$, the best result would also be approximately
$50\%$ better. However, his effect almost vanishes already at  $|\dm{31}| \sim 1.5\cdot 10^{-3} \, \mathrm{eV}^2$. 
Thus, we conclude that
there does not seem to be much improvement by fine tuning the combination of experiments for the 
reduction of the risk -- given all constraints on baselines and off-axis angles.

\section{Summary and conclusions}
\label{sec:summary}

In this paper, we have first presented an analysis of the {\sf JHF-SK} and {\sf NuMI} superbeam experiments as they are proposed in their letters of intent. We demonstrated that both are optimized under similar conditions, \ie, for a similar value of $L/E$, in order to measure the leading atmospheric parameters and $\sin^2 2 \theta_{13}$ at the current atmospheric best-fit values. However, we have also shown that there are no real synergy effects in these two experiments,  \ie, effects which go beyond the simple addition of statistics. A combined fit of their data shows only the improvement of the results coming from the better statistics, which is not surprising since two similar experiments combined simply perform better than one. To demonstrate this, we have plotted their individual results at double luminosities and we have compared them to their combined fit. Their combined results for $\sin^2 2 \theta_{13}$ are approximately as good as the best of the individual ones at double luminosity, even as a function of $\Delta m_{31}^2$.  In addition, we have not found any sensitivity to the sign of $\Delta m_{31}^2$, neither in the individual experiments, nor in their combination. Moreover, there is hardly any sensitivity to CP violation in the LMA allowed region. Thus, we conclude that for the current design as specified in the LOIs one could run one experiment equally well twice as long as planned (or with a twice as big detector) instead of building a second experiment.

As alternatives to the originally considered setups, we have proposed to run either one or both of the two experiments partly or entirely with antineutrinos, or to build the {\sf NuMI} detector at a different baseline.  We especially considered two alternative baselines of $890 \, \mathrm{km}$ at an off-axis angle of 0.72$^\circ$ and $950 \, \mathrm{km}$ at an off-axis angle of 0.97$^\circ$, which both would be still possible under the constraint of the fixed {\sf NuMI} decay pipe. Possible detector sites for these baselines are Fort Frances, Ontario ($L=875 \, \mathrm{km}$) and Vermilion Bay, Ontario ($L=950 \, \mathrm{km}$).
We have found that, under such modified conditions, one could then either measure the sign of $\Delta m_{31}^2$ or (maximal) CP violation, depending on the actual value of $\Delta m_{21}^2$, by combining the initial stage {\sf JHF-SK} and {\sf NuMI} experiments, \ie, with running times of five years each and detector masses as proposed in the letters of intent. 
\begin{figure}[ht!]
\begin{center}
\includegraphics[width=9cm]{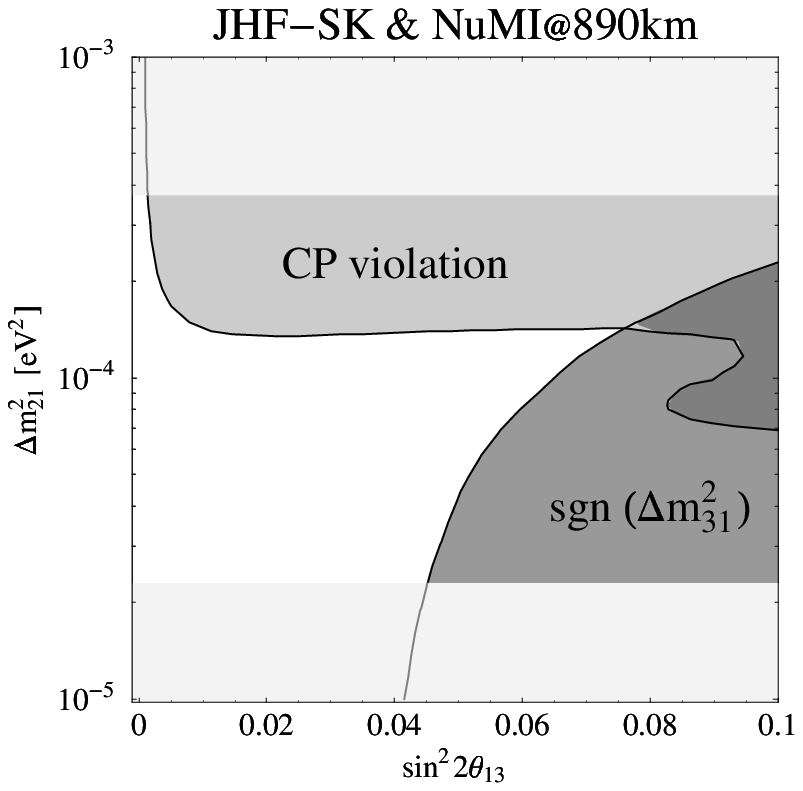}
\end{center}
\mycaption{\label{fig:exec1} The regions of sign of $\Delta m_{31}^2$ and CP violation sensitivity ({\sf JHF-SK} and {\sf NuMI} combined) in the $\sin^2 2 \theta_{13}$-$\Delta m_{21}^2$-plane (90\% confidence level, LMA values) for the $890\nu\nu$ (for the sign of $\Delta m_{31}^2$, neutrino running only) and $890cc$ (for CP violation, combined neutrino-antineutrino running) setups, where the LMA excluded region is shaded in gray. } 
\end{figure}
\fig~\ref{fig:exec1} shows the regions of sign of $\Delta m_{31}^2$ and CP violation sensitivity in the $\sin^2 2 \theta_{13}$-$\Delta m_{21}^2$-plane at the 90\% confidence level for the $890\nu\nu$ (sign of $\Delta m_{31}^2$) and $890cc$ (CP violation) setups as defined in \Tab~\ref{tab:scenarios}. One can easily see that, with these initial stage experiments, a simultaneous measurement of both quantities is hard to achieve. Thus, the strategy on what one wants to measure will strongly depend on the KamLAND result.

For the sign of $\Delta m_{31}^2$ sensitivity, the $890 \, \mathrm{km}$ and $950 \, \mathrm{km}$ options for the {\sf NuMI} baseline with neutrino running only at both {\sf JHF-SK} and {\sf NuMI} turn out to 
perform very well. Above $\Delta m_{21}^2 \sim 4.5 \cdot 10^{-5} \, \mathrm{eV}^2$ the $890 \, \mathrm{km}$ option does better (higher $\Delta m_{21}^2$ reach), below $\Delta m_{21}^2 \sim 4.5 \cdot 10^{-5} \, \mathrm{eV}^2$ the $950 \, \mathrm{km}$ does better (better $\sin^2 2 \theta_{13}$ reach).
\begin{figure}[ht!]
\begin{center}
\includegraphics[width=10cm]{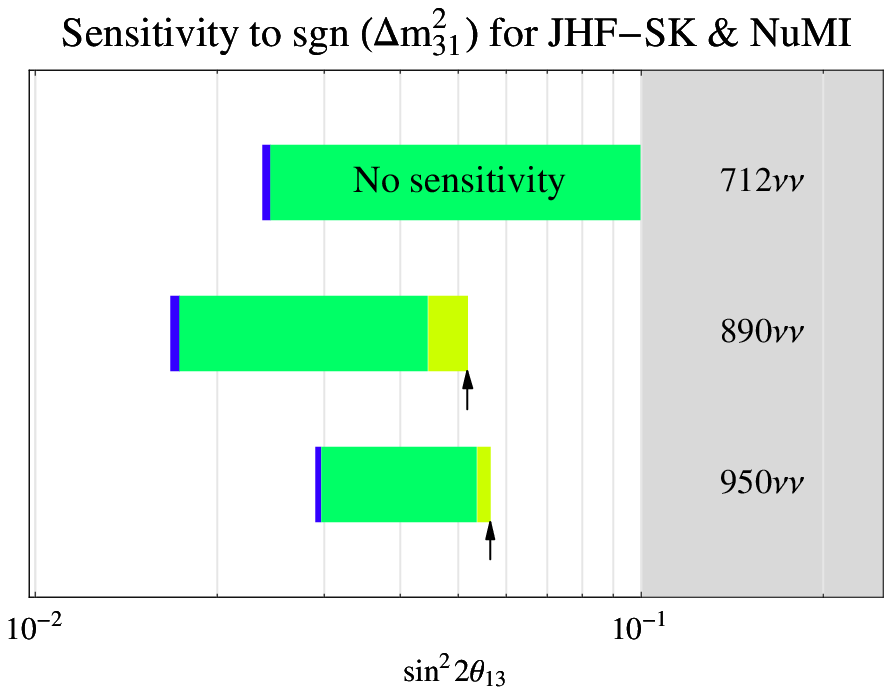}
\end{center}
\mycaption{\label{fig:exec2} The sensitivity limits to the sign of $\Delta m_{31}^2$ in the direction of $\sin^2 2 \theta_{13}$ ($90 \%$ confidence level, LMA values). The bars are shown for the combined {\sf JHF-SK} and {\sf NuMI} experiments for a selected set of the alternative options and the original combination (first bar), where the labels are explained in \Tab~\ref{tab:scenarios}. The left edges of the bars correspond to the sensitivity limits from statistics only. The  right edges of the bars correspond to the real sensitivity limits after switching on systematics (blue), correlations (green), and degeneracies (yellow) successively from the left to the right.} 
\end{figure}
\fig~\ref{fig:exec2} shows the sign of $\Delta m_{31}^2$ sensitivity limits at the LMA best-fit point for the original setup and the best of the alternative setups, including the influence of systematics, correlations, and degeneracies. In this figure, the left edges of the bars correspond to the sensitivity limits without taking into account the mentioned error sources, while successively switching on systematics, correlations, and degeneracies shifts the sensitivity limit to the right. Since it is very difficult to define the difference between correlations and degeneracies in this case, we fix $\deltacp=0$ at the border of the second and third bars. In either case, the right edges correspond to the real sensitivity limits. Thus, we do not have sensitivity to the sign of $\Delta m_{31}^2$ with the original setup, but we do have sensitivity at longer baselines. Which of the two proposed alternative baselines performs better depends on the exact value of $\Delta m_{21}^2$. However, from statistics only, the $890 \, \mathrm{km}$ option is much better.

For the sensitivity to CP violation, the $712 \, \mathrm{km}$ and $890 \, \mathrm{km}$ baseline options for {\sf NuMI} are performing very well -- provided that $\Delta m_{21}^2$ is large enough and there is a substantial fraction of antineutrino running at {\sf NuMI} (for a fixed total running time). Especially, running {\sf JHF-SK} with neutrinos and {\sf NuMI} with antineutrinos or running both experiments with neutrinos and antineutrinos with almost equal total numbers of neutrino and antineutrino events in each experiment provides very good results.
\begin{figure}[ht!]
\begin{center}
\includegraphics[width=10cm]{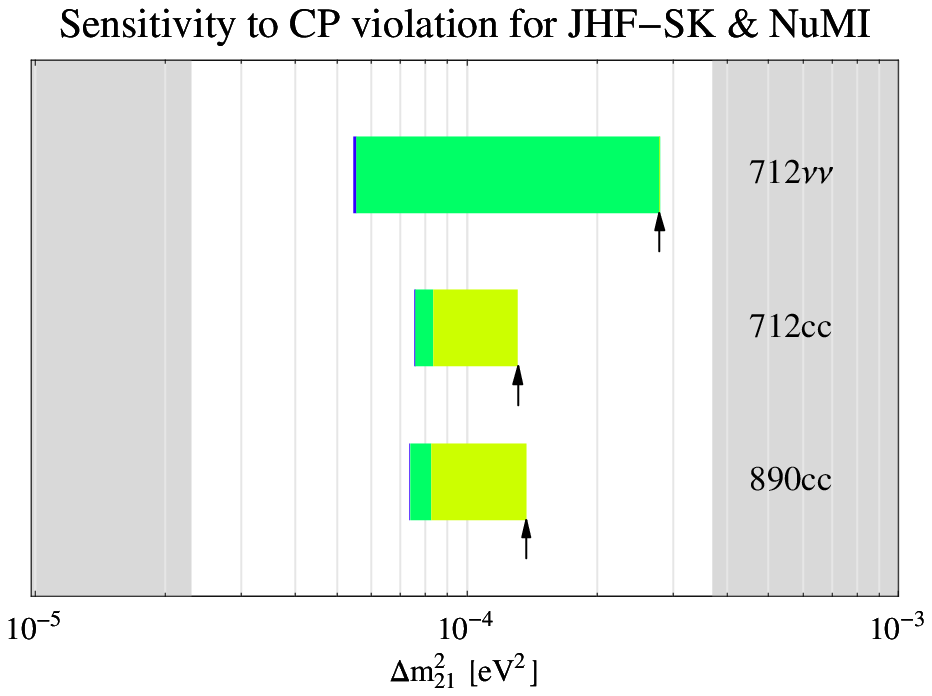}
\end{center}
\mycaption{\label{fig:exec3} The sensitivity limits to (maximal) CP violation plotted in the direction of $\Delta m_{21}^2$ at the fixed value of $\sin^2 2 \theta_{13}=0.03$  ($90 \%$ confidence level, LMA values).  The bars are shown for the combined {\sf JHF-SK} and {\sf NuMI} experiments for a selected set of the alternative options and the original combination (first bar), where the labels are explained in \Tab~\ref{tab:scenarios}. The left edges of the bars correspond to the sensitivity limits from statistics only. The  right edges of the bars correspond to the real sensitivity limits after switching on systematics (blue), correlations (green), and degeneracies (yellow) successively from the left to the right. The gray-shaded are marks the region of the LMA excluded values.} 
\end{figure}
\fig~\ref{fig:exec3} shows the sensitivity limits to maximal CP violation for a fixed value of $\sin^2 2 \theta_{13}=0.03$ for the therein specified combinations and experiments, where again the systematics, correlations, an degeneracies are switched on from the left to the right. The right edges of the bars correspond thus to the final sensitivity limits. Especially the correlation of $\deltacp$  with $\theta_{13}$ can be resolved very well in this case with the alternative options. The remaining dominant error source comes then from the $\mathrm{sgn} (\Delta m_{31}^2)$ degeneracy.

Summarizing these result, we find for the combination of the modified initial stage {\sf JHF-SK} and {\sf NuMI} experiments synergy effects, which means that they have the potential to do new physics for any value of $\Delta m_{21}^2$ in the LMA regime:
\begin{description}
\item[For $\Delta m_{21}^2 \gtrsim 1.0 \cdot 10^{-4} \, \mathrm{eV}^2$] one could be sensitive to (maximal) CP violation with a substantial fraction of antineutrino running at least in the {\sf NuMI} experiment, or in both.
\item[For $\Delta m_{21}^2 \sim 4.5 \cdot 10^{-5} \, \mathrm{eV}^2 - \, 1.0 \cdot 10^{-4} \, \mathrm{eV}^2$] one would be optimally sensitive towards the sign of $\Delta m_{31}^2$ with a {\sf NuMI} baseline of $\sim 890 \, \mathrm{km}$ and running both {\sf JHF-SK} and {\sf NuMI} with neutrinos only.
\item[For $\Delta m_{21}^2 \lesssim 4.5 \cdot 10^{-5} \, \mathrm{eV}^2$] one would be optimally sensitive towards the sign of $\Delta m_{31}^2$ with a {\sf NuMI} baseline of $\sim 950 \, \mathrm{km}$ and running both {\sf JHF-SK} and {\sf NuMI} with neutrinos only.
\end{description}
Especially, the options with a {\sf NuMI} baseline of $L=890 \, \mathrm{km}$ seems to be a good compromise independent of the forthcoming exact KamLAND results.

The proposed modifications for one or both of the experiments mean that the individual experiments are no longer optimized for the best-fit point by themselves anymore. They could therefore in many cases, as standalone experiments, not compete anymore with respect to the leading oscillation parameters. However, we have seen that the combined $\sin^2 2 \theta_{13}$ sensitivity limits, which in our definition do not depend on the sign of $\Delta m_{31}^2$, could even be
better by up to $20\%$ and in the worst case the reduction would only be 
$10\%$. For the atmospheric parameters those modifications should not affect 
the result since any of the individual setups is systematics limited for this
measurement. Thus, we conclude that it is possible to obtain exciting synergies for the combined {\sf JHF-SK} and {\sf NuMI} experiments if they are optimized together. Depending on $\Delta m_{21}^2$, the combination could then measure either the sign of $\Delta m_{31}^2$ or leptonic CP violation already within this decade.

%%%%%%%%%%%%%%%%%%%%%%%%%%%%%%%%%%%%%%%%%%%%%%%%%%%%%%%%%%%%%%%%%%%%
%%%%                      Acknowledgments                      %%%%
%%%%%%%%%%%%%%%%%%%%%%%%%%%%%%%%%%%%%%%%%%%%%%%%%%%%%%%%%%%%%%%%%%%%

\vspace*{7mm}
\subsubsection*{Acknowledgments}
We wish to thank Deborah~A.~Harris for valuable information on the
{\sf NuMI} experiment.

%%%%%%%%%%%%%%%%%%%%%%%%%%%%%%%%%%%%%%%%%%%%%%%%%%%%%%%%%%%%%%%%%%%%%
%%%%                       References                            %%%%
%%%%%%%%%%%%%%%%%%%%%%%%%%%%%%%%%%%%%%%%%%%%%%%%%%%%%%%%%%%%%%%%%%%%%

\newpage
\begin{appendix}
\section*{Appendix: Experiment Description}
\label{app:a}

For the simulation of the experiments, we use the same techniques and
notation as in \Ref~\cite{Huber:2002mx}. Here we therefore only give a short
description of each of the experiments and the numbers used in our
calculation.

\subsection*{{\sf JHF-SK}}
The Super-Kamiokande detector has an excellent capability to 
separate muons and electrons. Furthermore,
it provides a very accurate measurement of the charged lepton 
momentum~\cite{Itow:2001ee}. However it is completely lacking the measurement
of the hadronic fraction of a neutrino event. Therefore, the energy can only
be reconstructed for the QE (quasi-elastic scattering) events. 
For this sort of events,  the energy resolution is dominated by
the Fermi motion of the nucleons, which induces a constant width of 
$80-100\,\mathrm{MeV}$~\cite{Itow:2001ee,NAKAYA}. In order to incorporate
spectral information in our analysis, we use the spectrum of the QE
events with a free normalization and the total rate of all CC (charged current) events.
The free normalization of the spectrum is necessary to avoid double 
counting the QE events. For the energy resolution, we use a constant width 
of $\sigma_E=85\,\mathrm{MeV}$. The energy window for the analysis is 
$0.4-1.2\,\mathrm{GeV}$.
 The efficiencies and background fractions
are given in \Tab~\ref{tab:jhf} and taken from \Ref~\cite{Itow:2001ee}.
\begin{center}
\begin{table}[ht!] \begin{center} \begin{tabular}[h]{|l|lll|} \hline
\multicolumn{4}{|l|}{Disappearance}\\
\hline
{\small Signal}&$0.9\otimes(\ruu)_\mathrm{QE}$&&\\
&&&\\
{\small Background}&$0.0056\otimes(\rux)_\mathrm{NC}$&&\\
\hline
\multicolumn{4}{|l|}{Appearance}\\
\hline
{\small Signal}&$0.505\otimes(\rue)_\mathrm{CC}$&&\\
&&&\\
{\small
Background}&$0.0056\otimes(\rux)_\mathrm{NC}$&$3.3\cdot10^{-4}\otimes(\ruu)_\mathrm{
CC}$&\\
{\small Beam
background}&$0.505\otimes(\ree)_\mathrm{CC}$&$0.505\otimes(\reeb)_\mathrm{CC}$&\\
\hline
\end{tabular}
\mycaption{\label{tab:jhf} The efficiencies for the signals and backgrounds of
the {\sf JHF-SK} experiment.} 
\end{center} 
\end{table}
\end{center}
We furthermore include a  normalization uncertainty of $5\%$, as well as a background
uncertainty of $5\%$. The results are however not sensitive to
reasonable variations of those uncertainties, since the statistical error is 
much larger.
The fluxes of $\nu_\mu$, $\nu_e$, and $\bar{\nu}_e$
are the ones corresponding to $2^\circ$ OA beam in~\cite{Itow:2001ee} and have
been provided as data file by \Ref~\cite{NAKAYA}. 

\subsection*{{\sf NuMI}}
The {\sf NuMI} detector used in our calculations is a low-Z calorimeter as described
in \Ref~\cite{Ayres:2002nm}. It can measure both the lepton momentum and the
hadronic energy deposition, albeit with different accuracies. The resolution
is expected to be very similar to the MINOS detector~\cite{MINOS}. Therefore we
use a $\sigma_E=0.15\cdot E$, which has been shown in \Ref~\cite{Huber:2002mx} to be a very 
good approximation to the MINOS resolution. 
 The energy window for the analysis is $1.6-2.8\,\mathrm{GeV}$ for the 
$0.72^\circ$ off-axis beam and $1.2-2.5\,\mathrm{GeV}$ for the 
$0.97^\circ$ off-axis beam.
The efficiencies and 
background fractions are given in \Tab~\ref{tab:numi} and are taken 
from \Ref~\cite{Ayres:2002nm} as far as they are given in there. The missing 
information was provided by \Ref~\cite{HARRIS}.
\begin{center}
\begin{table}[ht!] \begin{center} 
\begin{tabular}[h]{|l|lll|} \hline 
\multicolumn{4}{|l|}{Disappearance}\\
\hline
{\small Signal}&$0.9\otimes(\ruu)_\mathrm{CC}$&&\\
&&&\\
{\small Background}&$0.005\otimes(\rux)_\mathrm{NC}$&&\\
\hline
\multicolumn{4}{|l|}{Appearance}\\
\hline
{\small Signal}&$0.4\otimes(\rue)_\mathrm{CC}$&&\\
&&&\\
{\small Background}&$0.005\otimes(\rux)_\mathrm{NC}$&&\\
{\small Beam background}&$0.4\otimes(\ree)_\mathrm{CC}$&$0.4\otimes(\reeb)_\mathrm{CC}$&\\
\hline
\end{tabular}
\mycaption{\label{tab:numi} The efficiencies for the signals and backgrounds of
the {\sf NuMI} experiment.} 
\end{center} 
\end{table}
\end{center}
We also include a $5\%$ uncertainty on the signal and background normalizations. Here also the results are not sensitive to
reasonable variations of those uncertainties, since the statistical error is 
much larger.
The fluxes, we use, are given in \Ref~\cite{fluxes} and have been provided
as data file by \Ref~\cite{HARRIS}.

\end{appendix}


\newpage

\begin{thebibliography}{10}
\expandafter\ifx\csname bibnamefont\endcsname\relax
  \def\bibnamefont#1{#1}\fi
\expandafter\ifx\csname bibfnamefont\endcsname\relax
  \def\bibfnamefont#1{#1}\fi
\expandafter\ifx\csname url\endcsname\relax
  \def\url#1{\texttt{#1}}\fi
\expandafter\ifx\csname urlprefix\endcsname\relax\def\urlprefix{URL }\fi
\providecommand{\bibinfo}[2]{#2}
\providecommand{\eprint}[2][]{\url{#2}}

\bibitem{Toshito:2001dk}
\bibinfo{author}{\bibfnamefont{T.}~\bibnamefont{Toshito}}
  (\bibinfo{year}{Super-Kamiokande, 2001}),
  \eprint[http://arXiv.org/abs]{hep-ex/0105023}.

\bibitem{Ahmad:2002jz}
\bibinfo{author}{\bibfnamefont{Q.~R.} \bibnamefont{Ahmad}} \emph{et~al.}
  (\bibinfo{collaboration}{SNO}), \bibinfo{journal}{Phys. Rev. Lett.}
  \textbf{\bibinfo{volume}{89}}, \bibinfo{pages}{011301}
  (\bibinfo{year}{2002}), \eprint[http://arXiv.org/abs]{nucl-ex/0204008}.

\bibitem{Ahmad:2002ka}
\bibinfo{author}{\bibfnamefont{Q.~R.} \bibnamefont{Ahmad}} \emph{et~al.}
  (\bibinfo{collaboration}{SNO}), \bibinfo{journal}{Phys. Rev. Lett.}
  \textbf{\bibinfo{volume}{89}}, \bibinfo{pages}{011302}
  (\bibinfo{year}{2002}), \eprint[http://arXiv.org/abs]{nucl-ex/0204009}.

\bibitem{Barger:2002iv}
\bibinfo{author}{\bibfnamefont{V.}~\bibnamefont{Barger}},
  \bibinfo{author}{\bibfnamefont{D.}~\bibnamefont{Marfatia}},
  \bibinfo{author}{\bibfnamefont{K.}~\bibnamefont{Whisnant}}, \bibnamefont{and}
  \bibinfo{author}{\bibfnamefont{B.~P.} \bibnamefont{Wood}},
  \bibinfo{journal}{Phys. Lett.} \textbf{\bibinfo{volume}{B537}},
  \bibinfo{pages}{179} (\bibinfo{year}{2002}),
  \eprint[http://arXiv.org/abs]{hep-ph/0204253}.

\bibitem{Bandyopadhyay:2002xj}
\bibinfo{author}{\bibfnamefont{A.}~\bibnamefont{Bandyopadhyay}},
  \bibinfo{author}{\bibfnamefont{S.}~\bibnamefont{Choubey}},
  \bibinfo{author}{\bibfnamefont{S.}~\bibnamefont{Goswami}}, \bibnamefont{and}
  \bibinfo{author}{\bibfnamefont{D.~P.} \bibnamefont{Roy}},
  \bibinfo{journal}{Phys. Lett.} \textbf{\bibinfo{volume}{B540}},
  \bibinfo{pages}{14} (\bibinfo{year}{2002}),
  \eprint[http://arXiv.org/abs]{hep-ph/0204286}.

\bibitem{Bahcall:2002hv}
\bibinfo{author}{\bibfnamefont{J.~N.} \bibnamefont{Bahcall}},
  \bibinfo{author}{\bibfnamefont{M.~C.} \bibnamefont{Gonzalez-Garcia}},
  \bibnamefont{and}
  \bibinfo{author}{\bibfnamefont{C.}~\bibnamefont{Pena-Garay}},
  \bibinfo{journal}{JHEP} \textbf{\bibinfo{volume}{07}}, \bibinfo{pages}{054}
  (\bibinfo{year}{2002}), \eprint[http://arXiv.org/abs]{hep-ph/0204314}.

\bibitem{deHolanda:2002pp}
\bibinfo{author}{\bibfnamefont{P.~C.} \bibnamefont{de~Holanda}}
  \bibnamefont{and} \bibinfo{author}{\bibfnamefont{A.~Y.}
  \bibnamefont{Smirnov}}  (\bibinfo{year}{2002}),
  \eprint[http://arXiv.org/abs]{hep-ph/0205241}.

\bibitem{Apollonio:1999ae}
\bibinfo{author}{\bibfnamefont{M.}~\bibnamefont{Apollonio}} \emph{et~al.}
  (\bibinfo{collaboration}{Chooz Collab.}), \bibinfo{journal}{Phys. Lett.}
  \textbf{\bibinfo{volume}{B466}}, \bibinfo{pages}{415} (\bibinfo{year}{1999}),
  \eprint{hep-ex/9907037}.

\bibitem{Nakamura:2001tr}
\bibinfo{author}{\bibfnamefont{K.}~\bibnamefont{Nakamura}}
  (\bibinfo{collaboration}{K2K}), \bibinfo{journal}{Nucl. Phys. Proc. Suppl.}
  \textbf{\bibinfo{volume}{91}}, \bibinfo{pages}{203} (\bibinfo{year}{2001}).

\bibitem{Paolone:2001am}
\bibinfo{author}{\bibfnamefont{V.}~\bibnamefont{Paolone}},
  \bibinfo{journal}{Nucl. Phys. Proc. Suppl.} \textbf{\bibinfo{volume}{100}},
  \bibinfo{pages}{197} (\bibinfo{year}{2001}).

\bibitem{Ereditato:2001an}
\bibinfo{author}{\bibfnamefont{A.}~\bibnamefont{Ereditato}},
  \bibinfo{journal}{Nucl. Phys. Proc. Suppl.} \textbf{\bibinfo{volume}{100}},
  \bibinfo{pages}{200} (\bibinfo{year}{2001}).

\bibitem{Barger:2000nf}
\bibinfo{author}{\bibfnamefont{V.}~\bibnamefont{Barger}},
  \bibinfo{author}{\bibfnamefont{S.}~\bibnamefont{Geer}},
  \bibinfo{author}{\bibfnamefont{R.}~\bibnamefont{Raja}}, \bibnamefont{and}
  \bibinfo{author}{\bibfnamefont{K.}~\bibnamefont{Whisnant}},
  \bibinfo{journal}{Phys. Rev.} \textbf{\bibinfo{volume}{D63}},
  \bibinfo{pages}{113011} (\bibinfo{year}{2001}),
  \eprint{arXiv:hep-ph/0012017}.

\bibitem{Gomez-Cadenas:2001eu}
\bibinfo{author}{\bibfnamefont{J.~J.} \bibnamefont{Gomez-Cadenas}}
  \emph{et~al.} (\bibinfo{collaboration}{CERN working group on Super Beams})
  (\bibinfo{year}{2001}), \eprint[http://arXiv.org/abs]{hep-ph/0105297}.

\bibitem{Itow:2001ee}
\bibinfo{author}{\bibfnamefont{Y.}~\bibnamefont{Itow}} \emph{et~al.}
  (\bibinfo{year}{2001}), \eprint[http://arXiv.org/abs]{hep-ex/0106019}.

\bibitem{Minakata:2001qm}
\bibinfo{author}{\bibfnamefont{H.}~\bibnamefont{Minakata}} \bibnamefont{and}
  \bibinfo{author}{\bibfnamefont{H.}~\bibnamefont{Nunokawa}},
  \bibinfo{journal}{JHEP} \textbf{\bibinfo{volume}{10}}, \bibinfo{pages}{001}
  (\bibinfo{year}{2001}), \eprint[http://arXiv.org/abs]{hep-ph/0108085}.

\bibitem{Aoki:2001rc}
\bibinfo{author}{\bibfnamefont{M.}~\bibnamefont{Aoki}} \emph{et~al.}
  (\bibinfo{year}{2001}), \eprint[http://arXiv.org/abs]{hep-ph/0112338}.

\bibitem{Aoki:2002ks}
\bibinfo{author}{\bibfnamefont{M.}~\bibnamefont{Aoki}}  (\bibinfo{year}{2002}),
  \eprint[http://arXiv.org/abs]{hep-ph/0204008}.

\bibitem{Barenboim:2002zx}
\bibinfo{author}{\bibfnamefont{G.}~\bibnamefont{Barenboim}},
  \bibinfo{author}{\bibfnamefont{A.}~\bibnamefont{De~Gouvea}},
  \bibinfo{author}{\bibfnamefont{M.}~\bibnamefont{Szleper}}, \bibnamefont{and}
  \bibinfo{author}{\bibfnamefont{M.}~\bibnamefont{Velasco}}
  (\bibinfo{year}{2002}), \eprint[http://arXiv.org/abs]{hep-ph/0204208}.

\bibitem{Whisnant:2002fx}
\bibinfo{author}{\bibfnamefont{K.}~\bibnamefont{Whisnant}},
  \bibinfo{author}{\bibfnamefont{J.~M.} \bibnamefont{Yang}}, \bibnamefont{and}
  \bibinfo{author}{\bibfnamefont{B.-L.} \bibnamefont{Young}}
  (\bibinfo{year}{2002}), \eprint[http://arXiv.org/abs]{hep-ph/0208193}.

\bibitem{Aoki:2002ae}
\bibinfo{author}{\bibfnamefont{M.}~\bibnamefont{Aoki}},
  \bibinfo{author}{\bibfnamefont{K.}~\bibnamefont{Hagiwara}}, \bibnamefont{and}
  \bibinfo{author}{\bibfnamefont{N.}~\bibnamefont{Okamura}}
  (\bibinfo{year}{2002}), \eprint[http://arXiv.org/abs]{hep-ph/0208223}.

\bibitem{Barenboim:2002nv}
\bibinfo{author}{\bibfnamefont{G.}~\bibnamefont{Barenboim}} \bibnamefont{and}
  \bibinfo{author}{\bibfnamefont{A.}~\bibnamefont{de~Gouvea}}
  (\bibinfo{year}{2002}), \eprint[http://arXiv.org/abs]{hep-ph/0209117}.

\bibitem{Ayres:2002nm}
\bibinfo{author}{\bibfnamefont{D.}~\bibnamefont{Ayres}} \emph{et~al.}
  (\bibinfo{year}{2002}), \eprint[http://arXiv.org/abs]{hep-ex/0210005}.

\bibitem{DeRujula:1998hd}
\bibinfo{author}{\bibfnamefont{A.~D.} \bibnamefont{Rujula}},
  \bibinfo{author}{\bibfnamefont{M.~B.} \bibnamefont{Gavela}},
  \bibnamefont{and}
  \bibinfo{author}{\bibfnamefont{P.}~\bibnamefont{Hernandez}},
  \bibinfo{journal}{Nucl. Phys.} \textbf{\bibinfo{volume}{B547}},
  \bibinfo{pages}{21} (\bibinfo{year}{1999}), \eprint{hep-ph/9811390}.

\bibitem{Barger:1999jj}
\bibinfo{author}{\bibfnamefont{V.~D.} \bibnamefont{Barger}},
  \bibinfo{author}{\bibfnamefont{S.}~\bibnamefont{Geer}},
  \bibinfo{author}{\bibfnamefont{R.}~\bibnamefont{Raja}}, \bibnamefont{and}
  \bibinfo{author}{\bibfnamefont{K.}~\bibnamefont{Whisnant}},
  \bibinfo{journal}{Phys. Rev.} \textbf{\bibinfo{volume}{D62}},
  \bibinfo{pages}{013004} (\bibinfo{year}{2000}),
  \eprint[http://arXiv.org/abs]{hep-ph/9911524}.

\bibitem{FLPR}
\bibinfo{author}{\bibfnamefont{M.}~\bibnamefont{Freund}},
  \bibinfo{author}{\bibfnamefont{M.}~\bibnamefont{Lindner}},
  \bibinfo{author}{\bibfnamefont{S.~T.} \bibnamefont{Petcov}},
  \bibnamefont{and} \bibinfo{author}{\bibfnamefont{A.}~\bibnamefont{Romanino}},
  \bibinfo{journal}{Nucl. Phys.} \textbf{\bibinfo{volume}{B578}},
  \bibinfo{pages}{27} (\bibinfo{year}{2000}),
  \eprint[http://arXiv.org/abs]{hep-ph/9912457}.

\bibitem{CERVERA}
\bibinfo{author}{\bibfnamefont{A.}~\bibnamefont{Cervera}} \emph{et~al.},
  \bibinfo{journal}{Nucl. Phys.} \textbf{\bibinfo{volume}{B579}},
  \bibinfo{pages}{17} (\bibinfo{year}{2000}), \bibinfo{note}{erratum ibid.
  Nucl. Phys. {\bf B593}, 731 (2001)}, \eprint{hep-ph/0002108}.

\bibitem{Barger:2000yn}
\bibinfo{author}{\bibfnamefont{V.~D.} \bibnamefont{Barger}},
  \bibinfo{author}{\bibfnamefont{S.}~\bibnamefont{Geer}},
  \bibinfo{author}{\bibfnamefont{R.}~\bibnamefont{Raja}}, \bibnamefont{and}
  \bibinfo{author}{\bibfnamefont{K.}~\bibnamefont{Whisnant}},
  \bibinfo{journal}{Phys. Rev.} \textbf{\bibinfo{volume}{D62}},
  \bibinfo{pages}{073002} (\bibinfo{year}{2000}),
  \eprint[http://arXiv.org/abs]{hep-ph/0003184}.

\bibitem{FHL}
\bibinfo{author}{\bibfnamefont{M.}~\bibnamefont{Freund}},
  \bibinfo{author}{\bibfnamefont{P.}~\bibnamefont{Huber}}, \bibnamefont{and}
  \bibinfo{author}{\bibfnamefont{M.}~\bibnamefont{Lindner}},
  \bibinfo{journal}{Nucl. Phys.} \textbf{\bibinfo{volume}{B585}},
  \bibinfo{pages}{105} (\bibinfo{year}{2000}), \eprint{hep-ph/0004085}.

\bibitem{Albright:2000xi}
\bibinfo{author}{\bibfnamefont{C.}~\bibnamefont{Albright}} \emph{et~al.}
  (\bibinfo{year}{2000}), \eprint{hep-ex/0008064, {\rm and references
  therein}}.

\bibitem{Burguet-Castell:2001ez}
\bibinfo{author}{\bibfnamefont{J.}~\bibnamefont{Burguet-Castell}},
  \bibinfo{author}{\bibfnamefont{M.~B.} \bibnamefont{Gavela}},
  \bibinfo{author}{\bibfnamefont{J.~J.} \bibnamefont{Gomez-Cadenas}},
  \bibinfo{author}{\bibfnamefont{P.}~\bibnamefont{Hernandez}},
  \bibnamefont{and} \bibinfo{author}{\bibfnamefont{O.}~\bibnamefont{Mena}},
  \bibinfo{journal}{Nucl. Phys.} \textbf{\bibinfo{volume}{B608}},
  \bibinfo{pages}{301} (\bibinfo{year}{2001}),
  \eprint[http://arXiv.org/abs]{hep-ph/0103258}.

\bibitem{Freund:2001ui}
\bibinfo{author}{\bibfnamefont{M.}~\bibnamefont{Freund}},
  \bibinfo{author}{\bibfnamefont{P.}~\bibnamefont{Huber}}, \bibnamefont{and}
  \bibinfo{author}{\bibfnamefont{M.}~\bibnamefont{Lindner}},
  \bibinfo{journal}{Nucl. Phys.} \textbf{\bibinfo{volume}{B615}},
  \bibinfo{pages}{331} (\bibinfo{year}{2001}),
  \eprint[http://arXiv.org/abs]{hep-ph/0105071}.

\bibitem{Yasuda:2001ip}
\bibinfo{author}{\bibfnamefont{O.}~\bibnamefont{Yasuda}}
  (\bibinfo{year}{2001}), \bibinfo{note}{and references therein},
  \eprint[http://arXiv.org/abs]{hep-ph/0111172}.

\bibitem{Bueno:2001jd}
\bibinfo{author}{\bibfnamefont{A.}~\bibnamefont{Bueno}},
  \bibinfo{author}{\bibfnamefont{M.}~\bibnamefont{Campanelli}},
  \bibinfo{author}{\bibfnamefont{S.}~\bibnamefont{Navas-Concha}},
  \bibnamefont{and} \bibinfo{author}{\bibfnamefont{A.}~\bibnamefont{Rubbia}},
  \bibinfo{journal}{Nucl. Phys.} \textbf{\bibinfo{volume}{B631}},
  \bibinfo{pages}{239} (\bibinfo{year}{2002}),
  \eprint[http://arXiv.org/abs]{hep-ph/0112297}.

\bibitem{Yasuda:2002jk}
\bibinfo{author}{\bibfnamefont{O.}~\bibnamefont{Yasuda}}
  (\bibinfo{year}{2002}), \bibinfo{note}{and references therein},
  \eprint[http://arXiv.org/abs]{hep-ph/0203273}.

\bibitem{Apollonio:2002en}
\bibinfo{author}{\bibfnamefont{M.}~\bibnamefont{Apollonio}} \emph{et~al.}
  (\bibinfo{year}{2002}), \eprint[http://arXiv.org/abs]{hep-ph/0210192}.

\bibitem{Donini:2002rm}
\bibinfo{author}{\bibfnamefont{A.}~\bibnamefont{Donini}},
  \bibinfo{author}{\bibfnamefont{D.}~\bibnamefont{Meloni}}, \bibnamefont{and}
  \bibinfo{author}{\bibfnamefont{P.}~\bibnamefont{Migliozzi}},
  \bibinfo{journal}{Nucl. Phys.} \textbf{\bibinfo{volume}{B646}},
  \bibinfo{pages}{321} (\bibinfo{year}{2002}),
  \eprint[http://arXiv.org/abs]{hep-ph/0206034}.

\bibitem{Wolfenstein:1978ue}
\bibinfo{author}{\bibfnamefont{L.}~\bibnamefont{Wolfenstein}},
  \bibinfo{journal}{Phys. Rev.} \textbf{\bibinfo{volume}{D17}},
  \bibinfo{pages}{2369} (\bibinfo{year}{1978}).

\bibitem{Wolfenstein:1979ni}
\bibinfo{author}{\bibfnamefont{L.}~\bibnamefont{Wolfenstein}},
  \bibinfo{journal}{Phys. Rev.} \textbf{\bibinfo{volume}{D20}},
  \bibinfo{pages}{2634} (\bibinfo{year}{1979}).

\bibitem{Mikheev:1985gs}
\bibinfo{author}{\bibfnamefont{S.~P.} \bibnamefont{Mikheev}} \bibnamefont{and}
  \bibinfo{author}{\bibfnamefont{A.~Y.} \bibnamefont{Smirnov}},
  \bibinfo{journal}{Sov. J. Nucl. Phys.} \textbf{\bibinfo{volume}{42}},
  \bibinfo{pages}{913} (\bibinfo{year}{1985}).

\bibitem{Mikheev:1986wj}
\bibinfo{author}{\bibfnamefont{S.~P.} \bibnamefont{Mikheev}} \bibnamefont{and}
  \bibinfo{author}{\bibfnamefont{A.~Y.} \bibnamefont{Smirnov}},
  \bibinfo{journal}{Nuovo Cim.} \textbf{\bibinfo{volume}{C9}},
  \bibinfo{pages}{17} (\bibinfo{year}{1986}).

\bibitem{Barger:2001yr}
\bibinfo{author}{\bibfnamefont{V.}~\bibnamefont{Barger}},
  \bibinfo{author}{\bibfnamefont{D.}~\bibnamefont{Marfatia}}, \bibnamefont{and}
  \bibinfo{author}{\bibfnamefont{K.}~\bibnamefont{Whisnant}},
  \bibinfo{journal}{Phys. Rev.} \textbf{\bibinfo{volume}{D65}},
  \bibinfo{pages}{073023} (\bibinfo{year}{2002}),
  \eprint[http://arXiv.org/abs]{hep-ph/0112119}.

\bibitem{Barger:2002rr}
\bibinfo{author}{\bibfnamefont{V.}~\bibnamefont{Barger}},
  \bibinfo{author}{\bibfnamefont{D.}~\bibnamefont{Marfatia}}, \bibnamefont{and}
  \bibinfo{author}{\bibfnamefont{K.}~\bibnamefont{Whisnant}},
  \bibinfo{journal}{Phys. Rev.} \textbf{\bibinfo{volume}{D66}},
  \bibinfo{pages}{053007} (\bibinfo{year}{2002}),
  \eprint[http://arXiv.org/abs]{hep-ph/0206038}.

\bibitem{Minakata:2002qi}
\bibinfo{author}{\bibfnamefont{H.}~\bibnamefont{Minakata}},
  \bibinfo{author}{\bibfnamefont{H.}~\bibnamefont{Nunokawa}}, \bibnamefont{and}
  \bibinfo{author}{\bibfnamefont{S.}~\bibnamefont{Parke}}
  (\bibinfo{year}{2002}), \eprint[http://arXiv.org/abs]{hep-ph/0208163}.

\bibitem{Burguet-Castell:2002qx}
\bibinfo{author}{\bibfnamefont{J.}~\bibnamefont{Burguet-Castell}},
  \bibinfo{author}{\bibfnamefont{M.~B.} \bibnamefont{Gavela}},
  \bibinfo{author}{\bibfnamefont{J.~J.} \bibnamefont{Gomez-Cadenas}},
  \bibinfo{author}{\bibfnamefont{P.}~\bibnamefont{Hernandez}},
  \bibnamefont{and} \bibinfo{author}{\bibfnamefont{O.}~\bibnamefont{Mena}},
  \bibinfo{journal}{Nucl. Phys.} \textbf{\bibinfo{volume}{B646}},
  \bibinfo{pages}{301} (\bibinfo{year}{2002}),
  \eprint[http://arXiv.org/abs]{hep-ph/0207080}.

\bibitem{Barger:2002xk}
\bibinfo{author}{\bibfnamefont{V.}~\bibnamefont{Barger}},
  \bibinfo{author}{\bibfnamefont{D.}~\bibnamefont{Marfatia}}, \bibnamefont{and}
  \bibinfo{author}{\bibfnamefont{K.}~\bibnamefont{Whisnant}}
  (\bibinfo{year}{2002}), \eprint[http://arXiv.org/abs]{hep-ph/0210428}.

\bibitem{Minakata:2002jv}
\bibinfo{author}{\bibfnamefont{H.}~\bibnamefont{Minakata}},
  \bibinfo{author}{\bibfnamefont{H.}~\bibnamefont{Sugiyama}},
  \bibinfo{author}{\bibfnamefont{O.}~\bibnamefont{Yasuda}},
  \bibinfo{author}{\bibfnamefont{K.}~\bibnamefont{Inoue}}, \bibnamefont{and}
  \bibinfo{author}{\bibfnamefont{F.}~\bibnamefont{Suekane}}
  (\bibinfo{year}{2002}), \eprint[http://arXiv.org/abs]{hep-ph/0211111}.

\bibitem{Huber:2002mx}
\bibinfo{author}{\bibfnamefont{P.}~\bibnamefont{Huber}},
  \bibinfo{author}{\bibfnamefont{M.}~\bibnamefont{Lindner}}, \bibnamefont{and}
  \bibinfo{author}{\bibfnamefont{W.}~\bibnamefont{Winter}},
  \bibinfo{journal}{Nucl. Phys.} \textbf{\bibinfo{volume}{B645}},
  \bibinfo{pages}{3} (\bibinfo{year}{2002}),
  \eprint[http://arXiv.org/abs]{hep-ph/0204352}.

\bibitem{offaxis}
\bibinfo{author}{\bibfnamefont{D.}~\bibnamefont{Beavis}} \emph{et~al.},
  \emph{\bibinfo{title}{Proposal of BNL AGS E-889}}, \bibinfo{type}{Tech.
  Rep.}, \bibinfo{institution}{BNL} (\bibinfo{year}{1995}).

\bibitem{fluxes}
\bibinfo{author}{\bibfnamefont{M.~D.} \bibnamefont{Messier}},
  \emph{\bibinfo{title}{Basics of the numi off-axis beam}}, \bibinfo{note}{talk
  given at "New Initiatives for the NuMI Neutrino Beam", May 2002, Batavia,
  IL}.

\bibitem{MESSIER}
\bibinfo{author}{\bibfnamefont{M.~D.} \bibnamefont{Messier}},
  \bibinfo{note}{private communication}.

\bibitem{BARGER}
\bibinfo{author}{\bibfnamefont{V.}~\bibnamefont{Barger}},
  \bibinfo{author}{\bibfnamefont{D.}~\bibnamefont{Marfatia}}, \bibnamefont{and}
  \bibinfo{author}{\bibfnamefont{B.}~\bibnamefont{Wood}},
  \bibinfo{journal}{Phys. Lett.} \textbf{\bibinfo{volume}{B498}},
  \bibinfo{pages}{53} (\bibinfo{year}{2001}), \eprint{\hfill \\
  hep-ph/0011251}.

\bibitem{Gonzalez-Garcia:2001zy}
\bibinfo{author}{\bibfnamefont{M.~C.} \bibnamefont{Gonzalez-Garcia}}
  \bibnamefont{and}
  \bibinfo{author}{\bibfnamefont{C.}~\bibnamefont{Pe$\tilde{\mathrm{n}}$a-Gara%
y}}, \bibinfo{journal}{Phys. Lett.} \textbf{\bibinfo{volume}{B527}},
  \bibinfo{pages}{199} (\bibinfo{year}{2002}),
  \eprint[http://arXiv.org/abs]{hep-ph/0111432}.

\bibitem{Geller:2001ix}
\bibinfo{author}{\bibfnamefont{R.~J.} \bibnamefont{Geller}} \bibnamefont{and}
  \bibinfo{author}{\bibfnamefont{T.}~\bibnamefont{Hara}}
  (\bibinfo{year}{2001}), \eprint[http://arXiv.org/abs]{hep-ph/0111342}.

\bibitem{Ota:2000hf}
\bibinfo{author}{\bibfnamefont{T.}~\bibnamefont{Ota}} \bibnamefont{and}
  \bibinfo{author}{\bibfnamefont{J.}~\bibnamefont{Sato}},
  \bibinfo{journal}{Phys. Rev.} \textbf{\bibinfo{volume}{D63}},
  \bibinfo{pages}{093004} (\bibinfo{year}{2001}),
  \eprint[http://arXiv.org/abs]{hep-ph/0011234}.

\bibitem{Ota:2002fu}
\bibinfo{author}{\bibfnamefont{T.}~\bibnamefont{Ota}} \bibnamefont{and}
  \bibinfo{author}{\bibfnamefont{J.}~\bibnamefont{Sato}}
  (\bibinfo{year}{2002}), \eprint[http://arXiv.org/abs]{hep-ph/0211095}.

\bibitem{Shan:2001br}
\bibinfo{author}{\bibfnamefont{L.-Y.} \bibnamefont{Shan}},
  \bibinfo{author}{\bibfnamefont{B.-L.} \bibnamefont{Young}}, \bibnamefont{and}
  \bibinfo{author}{\bibfnamefont{X.-m.} \bibnamefont{Zhang}},
  \bibinfo{journal}{Phys. Rev.} \textbf{\bibinfo{volume}{D66}},
  \bibinfo{pages}{053012} (\bibinfo{year}{2002}),
  \eprint[http://arXiv.org/abs]{hep-ph/0110414}.

\bibitem{Jacobsson:2001zk}
\bibinfo{author}{\bibfnamefont{B.}~\bibnamefont{Jacobsson}},
  \bibinfo{author}{\bibfnamefont{T.}~\bibnamefont{Ohlsson}},
  \bibinfo{author}{\bibfnamefont{H.}~\bibnamefont{Snellman}}, \bibnamefont{and}
  \bibinfo{author}{\bibfnamefont{W.}~\bibnamefont{Winter}},
  \bibinfo{journal}{Phys. Lett.} \textbf{\bibinfo{volume}{B532}},
  \bibinfo{pages}{259} (\bibinfo{year}{2002}),
  \eprint[http://arXiv.org/abs]{hep-ph/0112138}.

\bibitem{Shan:2002px}
\bibinfo{author}{\bibfnamefont{L.-Y.} \bibnamefont{Shan}} \bibnamefont{and}
  \bibinfo{author}{\bibfnamefont{X.-M.} \bibnamefont{Zhang}},
  \bibinfo{journal}{Phys. Rev.} \textbf{\bibinfo{volume}{D65}},
  \bibinfo{pages}{113011} (\bibinfo{year}{2002}).

\bibitem{Jacobsson:2002nb}
\bibinfo{author}{\bibfnamefont{B.}~\bibnamefont{Jacobsson}},
  \bibinfo{author}{\bibfnamefont{T.}~\bibnamefont{Ohlsson}},
  \bibinfo{author}{\bibfnamefont{H.}~\bibnamefont{Snellman}}, \bibnamefont{and}
  \bibinfo{author}{\bibfnamefont{W.}~\bibnamefont{Winter}}
  (\bibinfo{year}{2002}), \eprint[http://arXiv.org/abs]{hep-ph/0209147}.

\bibitem{PDG}
\bibinfo{author}{\bibnamefont{{Particle Data Group, D.E. Groom {\it et al.}}}},
  \bibinfo{journal}{Eur. Phys. J. C} \textbf{\bibinfo{volume}{15}},
  \bibinfo{pages}{1} (\bibinfo{year}{2000}), \bibinfo{note}{\hfill \\ {\tt
  http://pdg.lbl.gov/}}.

\bibitem{FREUND}
\bibinfo{author}{\bibfnamefont{M.}~\bibnamefont{Freund}},
  \bibinfo{journal}{Phys. Rev.} \textbf{\bibinfo{volume}{D64}},
  \bibinfo{pages}{053003} (\bibinfo{year}{2001}),
  \eprint[http://arXiv.org/abs]{hep-ph/0103300}.

\bibitem{Gonzalez-Garcia:2002mu}
\bibinfo{author}{\bibfnamefont{M.~C.} \bibnamefont{Gonzalez-Garcia}}
  \bibnamefont{and} \bibinfo{author}{\bibfnamefont{M.}~\bibnamefont{Maltoni}}
  (\bibinfo{year}{2002}), \eprint[http://arXiv.org/abs]{hep-ph/0202218}.

\bibitem{Maltoni:2000ib}
\bibinfo{author}{\bibfnamefont{M.}~\bibnamefont{Maltoni}},
  \bibinfo{journal}{Nucl. Phys. Proc. Suppl.} \textbf{\bibinfo{volume}{95}},
  \bibinfo{pages}{108} (\bibinfo{year}{2001}),
  \eprint[http://arXiv.org/abs]{hep-ph/0012158}.

\bibitem{HARRIS}
\bibinfo{author}{\bibfnamefont{D.~A.} \bibnamefont{Harris}},
  \bibinfo{note}{private communication}.

\bibitem{NAKAYA}
\bibinfo{author}{\bibfnamefont{T.}~\bibnamefont{Nakaya}},
  \bibinfo{note}{private communication}.

\bibitem{MINOS}
\bibinfo{author}{\bibfnamefont{J.}~\bibnamefont{Hylen}} \emph{et~al.}
  (\bibinfo{year}{NuMI Collaboration, 1997}),
  \bibinfo{note}{~FERMILAB-TM-2018}.

\end{thebibliography}
\end{document}